\documentclass[
superscriptaddress,
 amsmath,amssymb,
 aps,
 pra,
twocolumn
]{revtex4-2}

\usepackage{graphicx}
\usepackage{dcolumn}
\usepackage{bm}
\usepackage{lipsum}  
\usepackage{hyperref}
\hypersetup{
	colorlinks,
	linkcolor={blue},
	citecolor={blue},
	urlcolor={blue}
}

\usepackage{mathtools}
\usepackage[capitalise]{cleveref}
\usepackage[caption=false]{subfig}
\usepackage{braket}

\usepackage{algorithm, algorithmic}

\usepackage{xcolor}

\begin{document}

\title{Parallel window decoding enables scalable fault tolerant quantum computation}
\date{\today}

\author{Luka Skoric}
\affiliation{Riverlane, Cambridge, United Kingdom}
\author{Dan E. Browne}
\affiliation{Riverlane, Cambridge, United Kingdom}
\affiliation{Dept. of Physics and Astronomy, University College London, London, WC1E 6BT, UK}
\author{Kenton M. Barnes}
\affiliation{Riverlane, Cambridge, United Kingdom}
\author{Neil I. Gillespie}
\affiliation{Riverlane, Cambridge, United Kingdom}
\author{Earl T.~Campbell}
\affiliation{Riverlane, Cambridge, United Kingdom}
\affiliation{Dept. of Physics and Astronomy, University of Sheffield, Sheffield S3 7RH, UK}

\begin{abstract}

Large-scale quantum computers have the potential to hold computational capabilities beyond conventional computers for certain problems. However, the physical qubits within a quantum computer are prone to noise and decoherence, which must be corrected in order to perform reliable, fault-tolerant quantum computations. Quantum Error Correction (QEC) provides the path for realizing such computations. QEC continuously generates a continuous stream of data that decoders must process at the rate it is received, which can be as fast as 1 MHz in superconducting quantum computers. A little known fact of QEC is that if the decoder infrastructure cannot keep up, a data backlog problem~\cite{terhal2015quantum} is encountered and the quantum computer runs exponentially slower. Today’s leading approaches to quantum error correction are not scalable as existing decoders typically run slower as the problem size is increased, inevitably hitting the backlog problem. That is: the current leading proposal for fault-tolerant quantum computation is not scalable. Here, we show how to parallelize decoding to achieve almost arbitrary speed, removing this roadblock to scalability. Our parallelization requires some classical feed forward decisions to be delayed, leading to a slow-down of the logical clock speed. However, the slow-down is now only polynomial in code size, averting the exponential slowdown. We numerically demonstrate our parallel decoder for the surface code, showing no noticeable reduction in logical fidelity compared to previous decoders and demonstrating the parallelization speedup. 

\end{abstract}

\maketitle

\begin{figure}[t]
    \centering
    \includegraphics{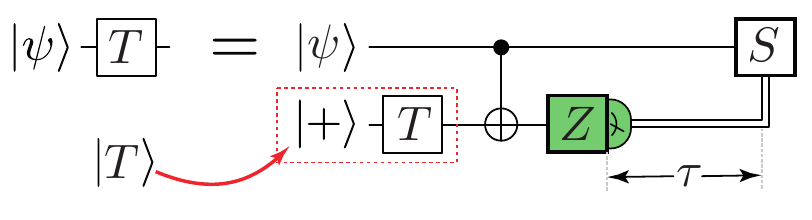}
    \caption{A gate-teleportation circuit to perform a $T$ gate using a magic state $\ket{T}:=T\ket{+}$, including a classically controlled $S$ gate depending on the measurement outcome.  In fault-tolerant implementations with logical qubits, the logical $Z$ measurement must be decoded before the $S$ correction can be correctly applied. This leads to a response time $\tau$ that is largely determined by the decoding time but also includes communication and control latency. }
    \label{fig:TeleportT}
\end{figure}


Quantum error correction (QEC) generates a stream of syndrome data to be decoded.  An offline decoder collects and stores all the syndrome data generated during a hardware run (often called a shot) and then performs decoding as a post-processing step. Offline decoding is sufficient for computations consisting solely of Clifford gates (e.g. CNOT and Hadamard gates).
However, fault-tolerant quantum computations must adapt in response to certain logical measurement results, which must be decoded to be reliable. For instance, when performing $T:=\mathrm{diag}(1, e^{i \pi/4})$ gates using teleportation and a magic state~\cite{bravyi2005universal,litinski2019game}, we must decide whether to apply a Clifford $S:=\mathrm{diag}(1, e^{i \pi/2})$ correction before performing the next non-Clifford operation (see \cref{fig:TeleportT}). This logic branching decision can only be reliably made after we decode the syndrome data from the $T$ gate teleportation \cite{divincenzo2007slow,terhal2015quantum,Chamberland2018faulttolerant}.   Therefore, online, or \textit{real-time}, decoding is necessary for useful quantum computation.   Classical computation occurs at finite speed, so online decoders will have some latency, but they need only react \textit{fast enough} to enable feed-forward and Clifford correction. 

How fast do decoders need to be?  A fundamental requirement was first noted by Terhal~\cite{terhal2015quantum} in her backlog argument
\begin{quote}
    ``Let $r_\mathrm{proc}$ be the rate (in bauds) at which syndrome bits are processed and $r_\mathrm{gen}$ be the rate at which these syndrome bits are generated. We can argue that if $r_{\mathrm{gen}}/r_{\mathrm{proc}} = f > 1$, a small initial backlog in processing syndrome data will lead to an \textit{exponential slow down} during the computation, \ldots  "
\end{quote}
Terhal proved that quantum algorithms with $T$-depth $k$  have a running time lower bounded by $c f^k$ when $f>1$ and $c$ is some constant.  Refs.~\cite{holmes2020nisq,chamberland2022techniques} provide more detailed reviews of this backlog argument. However, for all known decoders, as we scale the device decoding becomes more complex, the value of $f$ increases and inevitably we encounter the backlog problem.

Here we solve this problem, removing a fundamental roadblock to scalable fault-tolerant quantum computation. We propose parallelized window decoding that can be combined with any inner decoder that returns an (approximately) minimum weight solution, presenting results for minimum-weight perfect matching (MPWM)~\cite{dennis2002topological,fowler2009high,higgott2021pymatching} and union-find (UF)~\cite{delfosse2021almost,das2020scalable}. 

The previous leading idea was to modify decoders to work online was proposed by Dennis \textit{et al}~\cite{dennis2002topological}:
\begin{quote}``take action to remove only these long-lived defects, leaving those of more recent vintage to be dealt with in the next recovery step."
\end{quote}
Here \textit{defects} refer to observed changes in syndrome. Dennis \textit{et al} called this the \textit{overlapping recovery method} \cite{dennis2002topological,huang2021between}. Later, similar approaches were adopted for decoding classical LDPC codes~\cite{Iyengar2012sliding}, where this is known as \textit{sliding window decoding}. Roughly speaking, given a sequence of defects proceeding in time one decodes over some contiguous subset, or window. The decoder output gives only tentative error assignments, and from these only a subset --- those \emph{of an older vintage} --- are `committed’. Here, committing means making a final correction decision for potential error locations, with all corrections performed in software. One then slides the window up and the process repeats.

Sliding window decoding is inherently sequential. Let us consider a single code block (e.g.~a surface code patch) with each QEC round taking $\tau_{\mathrm{rd}}$ seconds. If each window is responsible for committing error corrections over $n_{\mathrm{com}}$ rounds of syndrome data, then it takes time $ n_{\mathrm{com}} \tau_{\mathrm{rd}}$ to generate all this data.  If the time to decode each window is $\tau_{\mathrm{W}}$, including any communication latency, then avoiding Terhal's backlog problem requires that $\tau_{\mathrm{W}}< n_{\mathrm{com}} \tau_{\mathrm{rd}}$. Since $\tau_{\mathrm{W}}$ typically grows superlinearly with the decoding volume, this leads to a hard upper bound on the achievable distance $d$.  For example, a distance $d$ surface code has $\tau_{\mathrm{W}}= \Omega( n_{\mathrm{com}} d^2)$ and therefore we are restricted to $d^2 \leq O(\tau_{\mathrm{rd}})$.     Scaling hardware based on a fixed device physics means $\tau_{\mathrm{rd}}$ is fixed. This imposes a hard limit on code distance.  The reader should pause to reflect how remarkable it is that the current leading proposal for fault-tolerant quantum computation is not scalable.

\begin{figure*}[t]
    \centering
    \includegraphics[width=500pt]{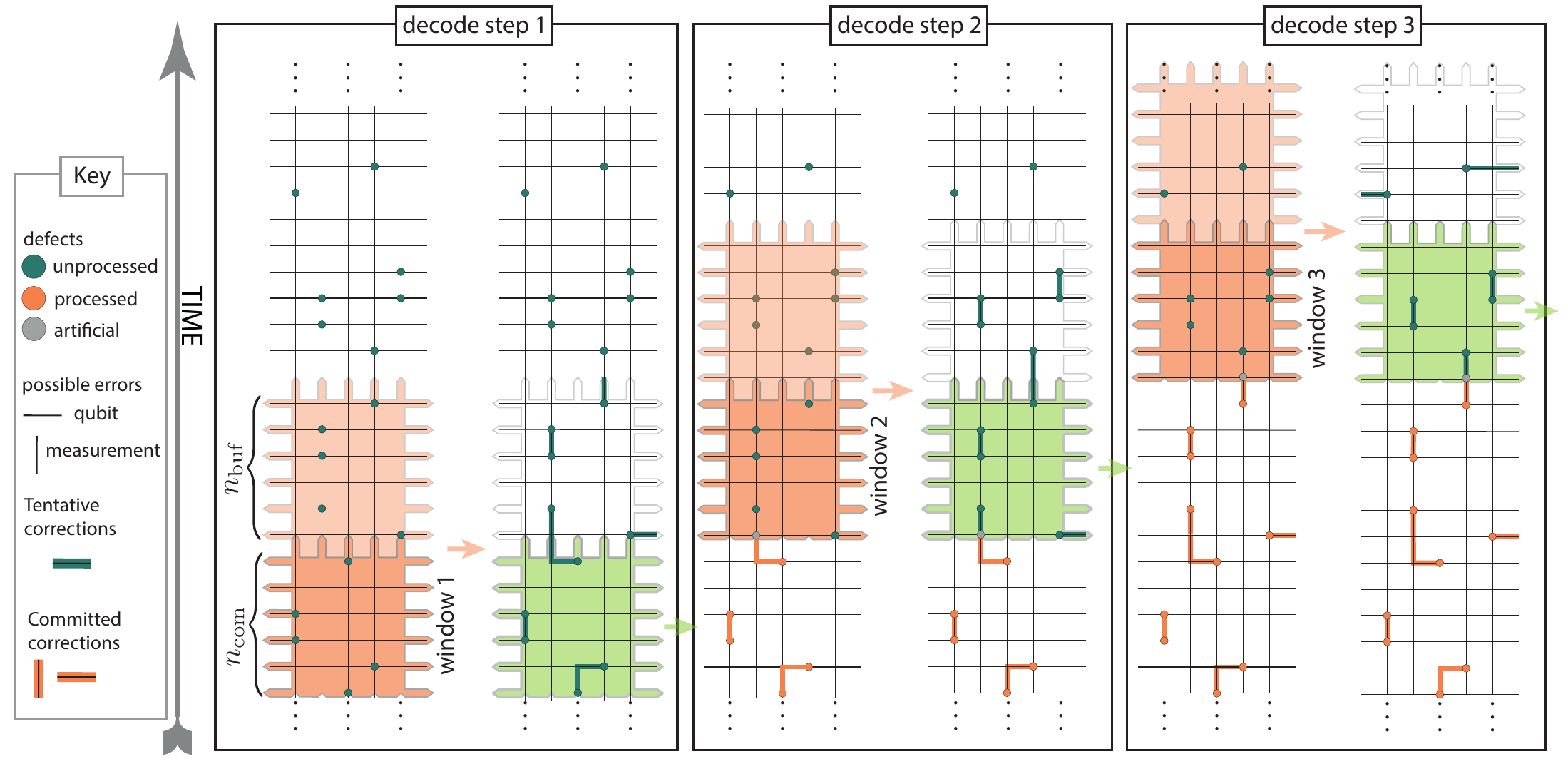}
    \caption{Sliding window decoding schematic. At each decoding step a number of syndrome rounds (window) is selected for decoding (orange region in left columns), and tentative corrections acquired. The corrections in the older part of the window (green region in right columns) are of high confidence and are committed to. The window is then moved up to the edge of the commit region and the process repeated.
        We decide to commit to the edges going from the commit region out of it, producing artificial defects defined by nodes outside of the region belonging to such an edge.
    }
    \label{fig:SlidingWindow}
\end{figure*}

As with sliding window decoding, our parallel window decoder breaks the problem up into sets of overlapping windows. Rather than solving these sequentially, some windows are decoded in parallel by adapting how overlapping windows are reconciled.  Through numeric simulations, we find that sliding, parallelized and global approaches differ in logical error rates by less than the error bars in our simulations. We show that, by scaling classical resources, parallel window can achieve almost arbitrarily high $r_{\mathrm{proc}}$ regardless of decoding time per window $\tau_{\mathrm{W}}$.
Furthermore, we show that while there is still an inherent latency determined by $\tau_{\mathrm{W}}$ leading to a slow-down of the logical clock speed, this is only linear in $\tau_{\mathrm{W}}$, rather than the exponential slow down resulting from Terhal's backlog argument.
We conclude with a discussion of the implications of this work for practical decoder requirements and extensions to a number of other decoding problems.  After making this work public, similar results were posted by the Alibaba team~\cite{tan2022scalable}.  The Alibaba numerics present the logical fidelity of the decoder, but do not include numerical results on decoding speed and improvements through increasing number of processors used. 

\section{Results and Discussion}

\subsection{Matching decoders}

Windowing techniques, both sliding and parallel, can be combined with most decoders acting internally on individual windows. We will refer to these as the ``inner decoders".  However, for brevity, in the main text we will describe the procedure for the case of matching decoders, such as MWPM and union-find.  A matching decoder is applicable when any error triggers either a pair of defects or a single defect.  For example, in the surface code $X$ errors lead to pairs of defects (when occurring in the bulk) or a single defect (when occurring at so-called \textit{rough} boundaries of the code).  To fully formulate a matching problem, all errors must lead to a pair of defects.  Therefore, errors triggering a single defect are connected to a virtual defect commonly called the boundary defect. We then have a graph where the vertices are potential defects (real or boundary) and edges represent potential errors.  Given an actual error configuration, we get a set of triggered defects and we can enforce that this is an even number by appropriately triggering the boundary defect.  A matching decoder takes as input this set of triggered defects and then outputs a subset of edges (representing a correction) that pair up the triggered defects.  Running a decoder on our entire defect data set at once (no windowing) will be referred to as global decoding, but global decoding is not compatible with the real-time feedback required for non-Clifford gates.

\begin{figure*}[t]
    \centering
    \includegraphics[width=440pt]{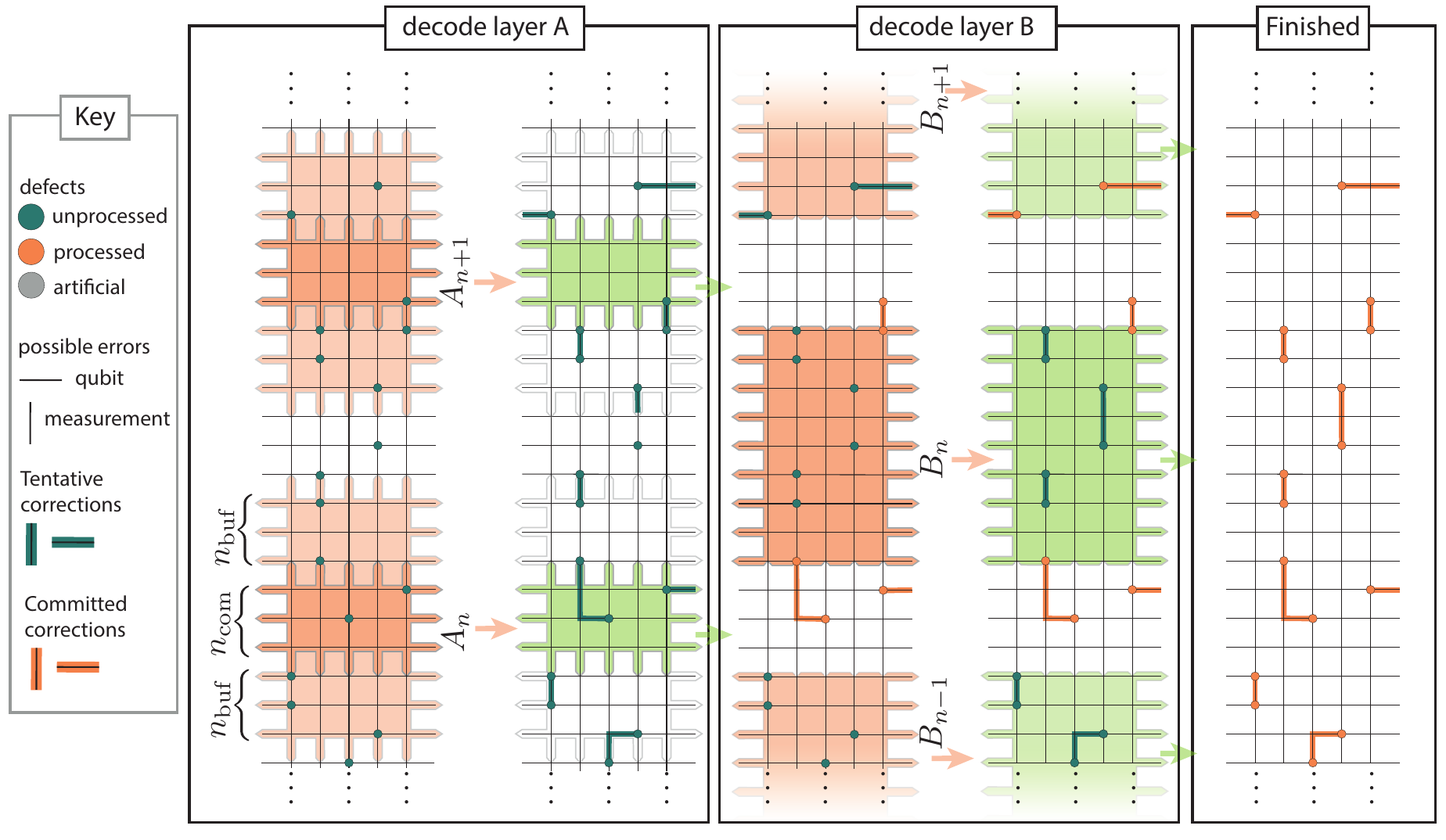}
    \caption{Parallel window decoding schematic for repetition code with extra spatial dimension added for surface codes. The decoding proceeds in two layers. In layer A, a number of non-overlapping windows is decoded in parallel. The high confidence corrections in the middle of each window are committed to, and the artificial defects passed on to layer B. Windows in layer B are fully committed to, resolving all the defects between the committed regions of layer A and completing the correction.}
    \label{fig:ParallelWindow}
\end{figure*}

\subsection{Sliding window decoding}

Instead of decoding a full history of syndrome data after the computation is complete,  sliding window decoding starts decoding the data in sequential steps while the algorithm is running. At each step, a subset (window) of $n_{\mathrm{W}}$ rounds of syndrome extraction is processed. The window correction graph is acquired by taking all the vertices and edges containing defects in the selected rounds. The measurement errors in the final round of a window only trigger a single defect within the window.
Therefore, all final round defects are additionally connected to the boundary defect, referred to as the rough top time boundary.

Following the overlapping recovery method~\cite{dennis2002topological,huang2021between}, a window can be divided into two regions: a \textit{commit region} consisting of the ``long-lived" defects in the first $n_{\mathrm{com}}$ rounds, and a \textit{buffer region} containing the last $n_{\mathrm{buf}}$ rounds ($n_{\mathrm{W}}=n_{\mathrm{com}}+n_{\mathrm{buf}}$).
An inner decoder (e.g.~MWPM or UF) outputs a subset of tentative correction edges within the window. Only the correction edges in the commit region are taken as final.
Sometimes, the chains of tentative correction edges will cross from the commit to the buffer region. Applying only the part of the chain in the commit region will introduce new defects, referred to as ``artificial defects'' along the boundary between the commit and buffer regions.  

The window is then moved up by $n_{\mathrm{com}}$ for the next decoding step that now includes the artificial defects along with the unresolved defects from the buffer region of the preceding step and new defects in the following rounds.  \cref{fig:SlidingWindow} illustrates sliding window for the simple example of a repetition code, naturally extending to surface codes by adding another spatial dimension.  Notice in \cref{fig:SlidingWindow} the creation of artificial defects where tentative corrections cross between commit and buffer regions. 

Due to these artificial defects, sliding window decoding (and also parallel window decoding, described below) requires an inner decoder which returns an approximately low weight correction, such as UF or MWPM. Decoders, such as those based on tensor network contractions, identify the optimal homology class (all error differing by stabilizers are in the same class) that contains a low-weight correction.  Once a homology class has been identified, we can always efficiently select a representative correction from the class but this could be a high weight correction (e.g. containing many stabilizer loops), leading to additional artificial defects at the boundary of the committed region, and then to logical errors when the next window is decoded. Therefore, additional modifications beyond those discussed in this work would be needed to use homology-based inner decoders.

Processing only a subset of the syndrome data at a time inevitably reduces the logical fidelity of the decoder. However, a logical fidelity close to that of the global decoder can be retained by making the unaccounted failure mechanisms negligible compared to the global failure rate. In particular, the error chains beginning in the committed region need to be unlikely (compared to the global failure rate) to span the buffer region and extend beyond the window. If the measurement and qubit error rates are comparable, to achieve this for distance $d$ codes, it suffices to make the buffer region of the same size $n_{\mathrm{buf}} = d$~\cite{dennis2002topological}. In the \cref{app:numerical_validation}, we demonstrate numerically that by choosing $n_{\mathrm{buf}} = n_{\mathrm{com}} = d$ there is no noticeable increase in logical error rate when applying the sliding window algorithm.

\subsection{Parallel window decoding}

Here we present our main innovation to overcome the backlog problem, which we call parallel window decoding. We illustrate the method in \cref{fig:ParallelWindow}. As in \cref{fig:SlidingWindow}, our illustration is for a repetition code example, naturally extending to a surface code, with further extensions discussed in \cref{sec:extensions}.


Parallel window decoding proceeds in two layers. First, we process a number of non-overlapping windows in decode layer A concurrently. As opposed to the sliding window approach, there are potentially unprocessed defects preceding the rounds in an A window. We thus need to include a buffer region both preceding and following the commit regions. Additionally, we set both time boundaries to be rough, connecting the first and last round of defects to the boundary node.
We set  $n_{\mathrm{buf}} = n_{\mathrm{com}} = w$, giving a total of $n_\mathrm{W} = 3w$ per window for some constant $w$. Using the same reasoning as with the sliding window we set $w=d$.  Note that in  \cref{fig:ParallelWindow} we use $w<d$ to keep the illustration compact.

Having committed to corrections in adjacent windows and computed the resulting artificial defects, in layer B we fill in the corrections in the rounds between the neighbouring A commit regions. For convenience, we separate A windows by $d$ rounds, so that B windows also have $n_\mathrm{W} = 3d$ rounds.
As the corrections preceding and succeeding the rounds in B windows have been resolved in layer A, the B windows have smooth time boundaries and do not require buffers.

\begin{figure*}[t]
    \centering
    \includegraphics[width=450pt]{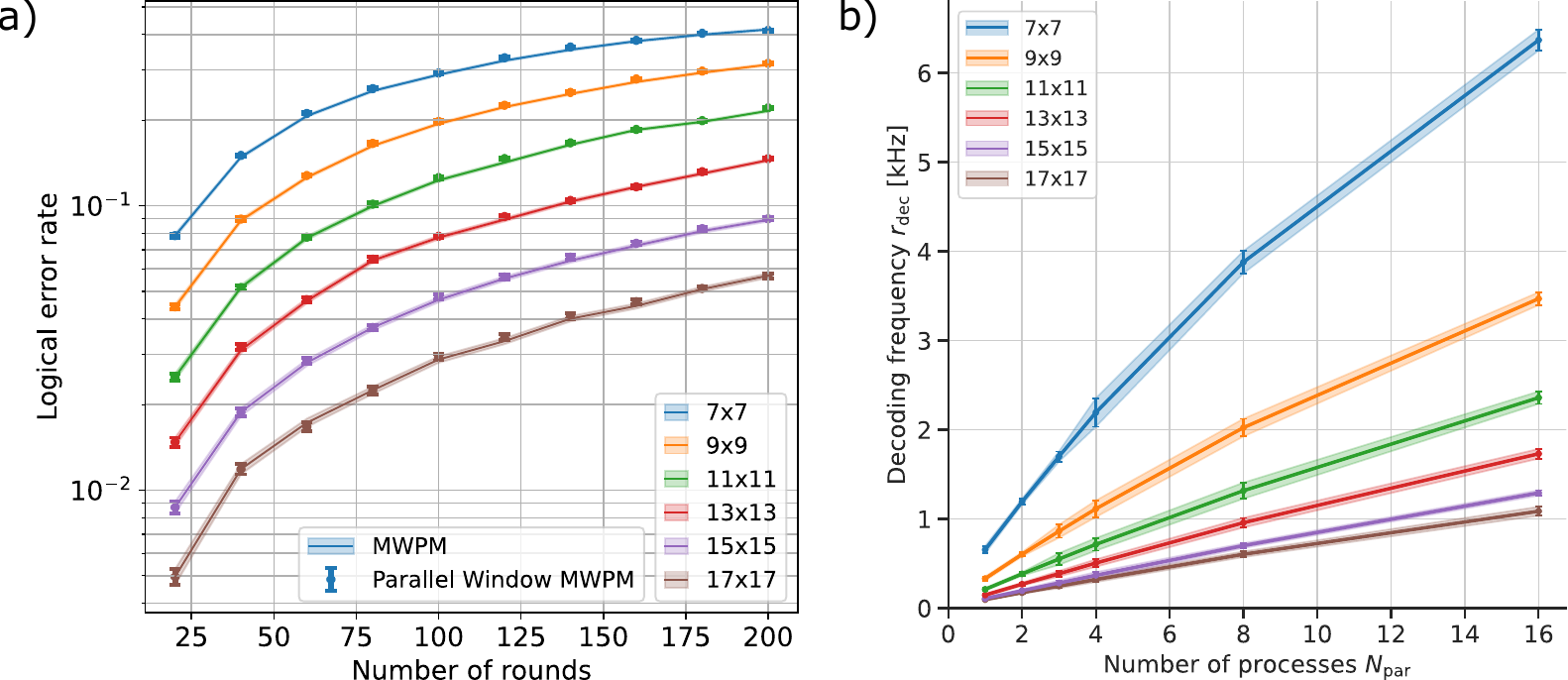}
    \caption{Logical error rate and decoding frequency on a rotated planar code using Minimum Weight Perfect Matching (MWPM) under phenomenological Pauli noise with 2\% physical error rate.
        (a) Logical error rates as a function of the number of rounds of syndrome extraction for different code sizes for both the global offline MWPM (shaded bands), and using the parallel window algorithm (points). The parallel window decoder has no numerically significant drop in logical fidelity compared to the global decoder.
        (b) The decoding frequency (number of rounds decoded per second) as a function of the number of decoding processes for the parallel window algorithm. The decoding frequency increases approximately linearly with the number of processes, achieving an order of magnitude faster decoding when using 16 processes. The sub-linearity most noticeable on small decoding problems is due to the parallelization overhead in the software implementation. Where the error bars are not visible, they are smaller than the marker size.  Here we plot the decoding frequency $r_{\mathrm{dec}}$, therefore the rate of syndrome processing is $r_{\mathrm{proc}} =  r_{\mathrm{dec}} (d^2-1) $. }
    \label{fig:MWPMparallelWindow}
\end{figure*}

Crucially, if the size of windows and the commit region in layer A are chosen appropriately, we expect no significant drop in logical fidelity compared to the global decoder. As with sliding windows, this is because each error chain of length $\leq d$ is guaranteed to be fully captured within one of the windows. In \cref{fig:MWPMparallelWindow}a we verify this by simulating the decoding process. We find that the logical error rates of rotated planar codes using the global MWPM and parallel window MWPM are within the numerical error of each other across a range of code sizes and number of measurement rounds. The same holds for UF-based decoders with data presented in the \cref{app:numerical_validation}.

This approach is highly parallelizable: as soon as the last round of window $A_n$ has been measured, the data can be given to a worker process to decode it. However, as the window $B_n$ requires the artificial defects generated by windows $A_n$ and $A_{n+1}$ adjacent to it (see \cref{fig:ParallelWindow}), it can only start once both processes have completed.
In the \cref{app:pipelining}, we sketch a schematic defining how the data pipelining could be implemented in an online parallel window decoder to achieve a high utilization of available decoding cores.

Assuming no parallelization overhead, the syndrome throughput will scale linearly with the number of parallel processes $N_\mathrm{par}$.
In this case, $N_\mathrm{par} n_{\mathrm{com}}$ rounds are committed to in layer A, and  $N_\mathrm{par} n_{\mathrm{W}}$ in layer B. Each round takes $\tau_{\mathrm{rd}}$ to acquire and the two layers of decoding take $2\tau_\mathrm{W}$.
To avoid the backlog problem, we need the acquisition time to be greater than the decoding time:
\begin{equation}
    N_\mathrm{par}(n_{\mathrm{com}} + n_{\mathrm{W}})\tau_{\mathrm{rd}} \ge  2\tau_{\mathrm{W}}.
\end{equation}
Therefore, the number of processes needs to be at least:
\begin{equation}
    \label{eqn:N_par}
    N_\mathrm{par} \ge  \frac{2\tau_{\mathrm{W}}}{(n_{\mathrm{com}} + n_{\mathrm{W}})\tau_{\mathrm{rd}}}  .
\end{equation}

In practice, the overhead of data communication among worker processes needs to be considered.
In the parallel window algorithm, each process only needs to receive defect data before it is started, and return the artificial defects and the overall effect of the committed correction on the logical operators (see \cref{app:pipelining}). Thus, we expect the data communication overhead to be negligible compared to the window decoding time. Indeed, in \cref{fig:MWPMparallelWindow}b we demonstrate this by simulating parallel window decoding in Python using MWPM as the inner decoder, showing how using $N_{\mathrm{par}}=16$ leads to greater than an order-of-magnitude increase in decoding speed.
Some sub-linearity can be seen due to parallelization overheads in software, particularly for low-distance codes where the decoding problem is relatively simple. In the \cref{app:numerical_validation}, we repeat these simulations using UF decoder where the overhead is more noticeable due to faster decoding of individual windows. However, hardware decoders such as FPGA (Field Programmable Gate Array) and ASIC (Application-Specific Integrated Circuit) are more suited to parallel data processing, allowing a large number of processes without being bottle-necked by the communication overheads (discussed further in \cref{app:pipelining}).  Lastly, even with some sub-linearity, the backlog can be averted provided as we really only need that arbitrary decoding speed is achieved with polynomial number of processors.   

\subsection{Resulting resource overheads}

While we can achieve almost arbitrarily high syndrome processing rates, there is still an inherent latency determined by the time to decode each window $\tau_{\mathrm{W}}$.
If $\tau_{\mathrm{W}}$ is large compared to the physical QEC round time $\tau_{\mathrm{rd}}$, we may slow down the logical clock of the quantum computer to compensate for this latency.  This slowdown is achieved simply by extending the delay time $\tau$ as shown in \cref{fig:TeleportT}.
If we pick $N_\mathrm{par}$ as described in \cref{eqn:N_par}, at every instance a block of $n_\mathrm{lag} = N_\mathrm{par}(n_{\mathrm{com}} + n_{\mathrm{W}})$ rounds are being decoded at once. The last round for which the full syndrome history has been decoded is therefore going to be $n_\mathrm{lag}$ rounds behind the most recently measured syndrome data. Therefore, we can set the response time after each $T$-gate (as defined in \cref{fig:TeleportT}) to
\begin{equation} \label{eq_response}
    \tau = n_\mathrm{lag}\tau_\mathrm{rd} = N_\mathrm{par}(n_{\mathrm{com}} + n_{\mathrm{W}}) \tau_\mathrm{rd}
\end{equation}
However, combining \cref{eqn:N_par}  and \cref{eq_response} the total response time is  $\approx 2 \tau_W$.  That is, for an algorithm with $k$ layers of $T$ gates, the total response time is $\tau k \approx 2 k \tau_W$.  This is in stark contrast to the exponential in $k$ response time observed by Terhal~\cite{terhal2015quantum}.  Furthermore, using an efficient decoder for each window, the average window decode time $\tau_W$ scales polynomially with code size $d$, so $\tau_W=O(d^{\alpha})$ for some constant $\alpha$.  Since code size is poly-logarithmic in algorithm depth $k$ and width $W$, $d=O(\log(k W)^\beta)$ for some constant $\beta$.  The response time per layer of $T$-gates is a poly-logarithmic factor so $\tau = O(\log(k W)^{\alpha\beta})$.  Strictly speaking, this additional overhead increases the decoding volume $k W$ by a logarithmic factor, but overall still gives a poly-logarithmic complexity.

We define logical clock time as how long it takes to execute one logical non-Clifford gate.  Using lattice surgery to perform $T$-teleportation, and assuming no bias between measurement and physical errors, takes $d \tau_{\mathrm{rd}}$ time for lattice surgery and $\tau$ response time. This gives a logical clock time of $\tau_{\mathrm{clock}}:=d \tau_{\mathrm{rd}} + \tau$.
Alternatively, this time overhead can be converted into a qubit overhead by moving Clifford corrections into an auxiliary portion of the quantum computer~\cite{austinfowlertimeoptimalpaper}, for example using auto-corrected $T$-gate teleportation~\cite{litinski2019game,gidney2019flexible}. In algorithm resource analysis, a common assumption is that $T$ gates are performed sequentially ~\cite{litinski2019game,berry2019qubitization,kivlichan2020improved,campbell2021early,chamberland2022universal,lee2021even,von2021quantum,blunt2022perspective,chamberland2022building} as then only a few magic-state factories are needed to keep pace.   Auto-correction gadgets enable us to perform the next $T$-gate before the response time has elapsed.  The price is that an auxiliary logical qubit must instead be preserved for time $\tau$, after which it is measured in a Pauli basis depending on the outcome of the decoding problem. Therefore, instead of a time overhead we can add $\lceil \tau / d\tau_{\mathrm{rd}} \rceil$ auxiliary logical qubits.  If we have an algorithm with 100 logical qubits and $\tau_{\mathrm{clock}}= 10 d \tau_{\mathrm{rd}}$, then: \textit{without} auto-correction we incur a $10\times$time cost; and \textit{with} auto-correction we instead require 9 auxiliary logicals qubits and so a $1.09 \times$qubit cost.  Under these common algorithm resource assumptions, we find seemingly large time overheads from parallel window decoding can be exchanged for modest qubit overheads. Indeed, the auto-correction strategies trade time for space resource, but the overall space-time volume is preferable under these resource estimation assumptions ($1.09\times$ instead of $10\times$).  Note that the additional space-time volume required for magic state distillation will depend only on the number of magic states produced and not on whether we use auto-corrected teleportation.

\subsection{Extensions}
\label{sec:extensions}

Error mechanisms (e.g.~$Y$ errors in the bulk of the surface code) sometimes trigger more than a pair of defects, but reasonable heuristics can often be used to approximately \textit{decorrelate} these errors to produce a graphical decoding problem. This decorrelation works well for the surface code.  However, many codes cannot be decorrelated and require a non-matching decoder.  Even when decorrelation approximations are possible,  logical fidelities can be improved by using a non-matching decoder that accounts for this correlation information~\cite{darmawan2017tensor,panteleev2021degenerate,roffe2020decoding,higgott2022fragile}.  Extensions of parallel window decoding to non-matching inner decoders are outlined in \cref{App:nonmatching}.

By judicious choice of window shapes and boundaries, one could consider 3D-shaped windows that divide the decoding problem in both space and time directions.  Similarly, we can construct 3D-shaped windows for parallel execution with only a constant number of layers.  When slicing in the time direction we only needed 2 layers of windows, but when constraining window size in $D$ dimensions a $D+1$ layer construction is possible, with the minimum number of layers being determined by the colorability of some tiling (see \cref{App:higherDim} for details).  When performing computation by lattice surgery, during merge operations the code temporally has an extended size~\cite{horsman2012surface,litinski2019game,chamberland2022universal,chamberland2022circuit}, and windowing in the spatial direction will become necessary to prevent the window decode time $\tau_{W}$ from significantly increasing. One may also wish to spatially window for a single logical qubit with windows smaller than the code distance since the decoder running time $\tau_W$ reduces with window size, and therefore the logical clock time may decrease (alternatively auto-correction qubit overhead may reduce).  But there are subtle tradeoffs.  Firstly, for windows of size $\omega<d$ in either the space or time direction, there may be adversarial failure mechanisms of weight $(\omega+1)/2 < (d+1)/2$ that are no longer correctly decoded.  One may speculate that this reduces the effective code distance to $\omega$.  However, in practice, percolation theory arguments~\cite{fawzi2018constant} show that for a distance $d$ code, the largest error clusters are typically of size $O(\mathrm{polylog}(d))$. This leaves open the possibility that windows of size $O(\mathrm{polylog}(d)) < \omega < d$ will suffice and be of practical value for stochastic (even if not adversarial) noise, though substantial further investigation is required.  We remark that this discussion assumes that measurement errors (that create vertical error chains) have a comparable probability as physical Pauli errors.  If there is a large measurement error bias, then we must appropriately scale the duration of lattice surgery operations and the vertical extent of our windows. 

\section{Conclusions}

Parallel window decoding avoids the exponential backlog growth that is unavoidable (for large enough computations) with sliding window decoders.
For many leading hardware platforms, such as superconducting devices, syndrome backlog can be a severe practical obstacle, even for modest code sizes.  In recent superconducting experiments a QEC round was performed every 1.1$\mu$s by Krinner \textit{et al.}~\cite{krinner2021realizing}  and every $921$ns by the Google Quantum AI team~\cite{acharya2022suppressing}. Our results are applicable to all hardware platforms, but the speed of superconducting quantum computers means these are amongst the most challenging systems for real-time decoding. Indeed, both aforementioned teams instead performed offline decoding, omitting a crucial aspect of scalable error correction.

To meet this challenge, improving the speed of decoders is currently an area of intense research.
For example,  LILLIPUT~\cite{das2021lilliput} is a recently proposed fast online sliding window decoder, implemented as an FPGA-based look-up table.  For $d\leq 5$ surface codes, the authors reported that a round of syndrome data could be processed every  $300$ns, fast enough even for superconducting qubits.
However, the memory requirements of lookup tables scale exponentially in qubit number, making this decoder impractical for all but the smallest code sizes. The UF decoder scales favourably, and
modelling of it on a dedicated microarchitecture~\cite{das2020scalable} suggested it would be fast enough for distance 11 surface codes.  However, the authors acknowledged ``further study is necessary to confirm the validity of our model in a real device".  Riverlane has recently released performance data, showing that real-time FPGA decoding should be possible on superconducting hardware with up-to distance 9 codes~\cite{riverlane2022whitepaper}. There have been other approaches to accelerating decoders.  A parallelized version of minimum weight perfect matching (MWPM) has been proposed~\cite{Fowler2013O1} but never implemented and its performance is unclear.  Adding a predecoding stage has also been identified as a way to further accelerate decoding and potentially boost  logical fidelity \cite{anwar2014fast,ueno2021qecool,meinerz2022scalable,chamberland2022techniques,paler2022pipelined,ueno2022neo}, but this has not been tested in an online setting. As such, even for modest code distance such as $d=11$, it is unclear whether conventional decoding approaches will be fast enough.


On the other hand, a parallel window decoder, as introduced here can achieve almost arbitrarily high decoding speed given enough classical resources and some (polynomially scaling) quantum resource overheads.  Therefore, this approach resolves both fundamental scalability issues and practical obstacles for hardware with rapid QEC cycle times.

\section{Methods}

All simulations were performed on an AMD EPYC 7742 processor. We used the PyMatching package \cite{higgott2021pymatching} to perform MWPM. For UF we used a custom Python implementation of the algorithm described in Ref. \cite{delfosse2021almost}.

In all experiments, phenomenological Pauli noise with physical error rate $p$ was used, meaning that there is a probability $p$ for a data error on every qubit at each round. Further, every syndrome measurement had an error with probability $p$.

To compute the timing for \cref{fig:MWPMparallelWindow}b and additional results in the \cref{app:numerical_validation}, we perform the decoding on $8(N_\mathrm{par}+1)d$ rounds to ensure a full two cycles of parallel decoding, averaging over 5000 repetitions. We assume initialisation and readout in the Z basis, meaning that the initial and final rounds of defects are smooth. Moreover, in parallel window decoding, we take the first round to always ``belong'' to layer A, and the first $2d$ rounds of the first window are committed to.
The last round belongs to a layer B if the total number of rounds $n_\mathrm{tot}$ satisfies $n_\mathrm{tot}\mod 4d \in (-d, d]$,  in which case the decoding is performed normally with the last B window potentially being of reduced size. Otherwise, the last window belongs to layer A and the commit region of the last window is from the bottom of the regular commit region to the last round.

\acknowledgements

This project made use of code co-developed with Adam Richardson and Joonas Majaniemi. We thank Kauser Johar for useful discussions.  We thank Steve Brierley and Jake Taylor for encouraging this research and related discussions.

\appendix

\begin{figure*}
    \centering
    \includegraphics[width=1.4 \columnwidth]{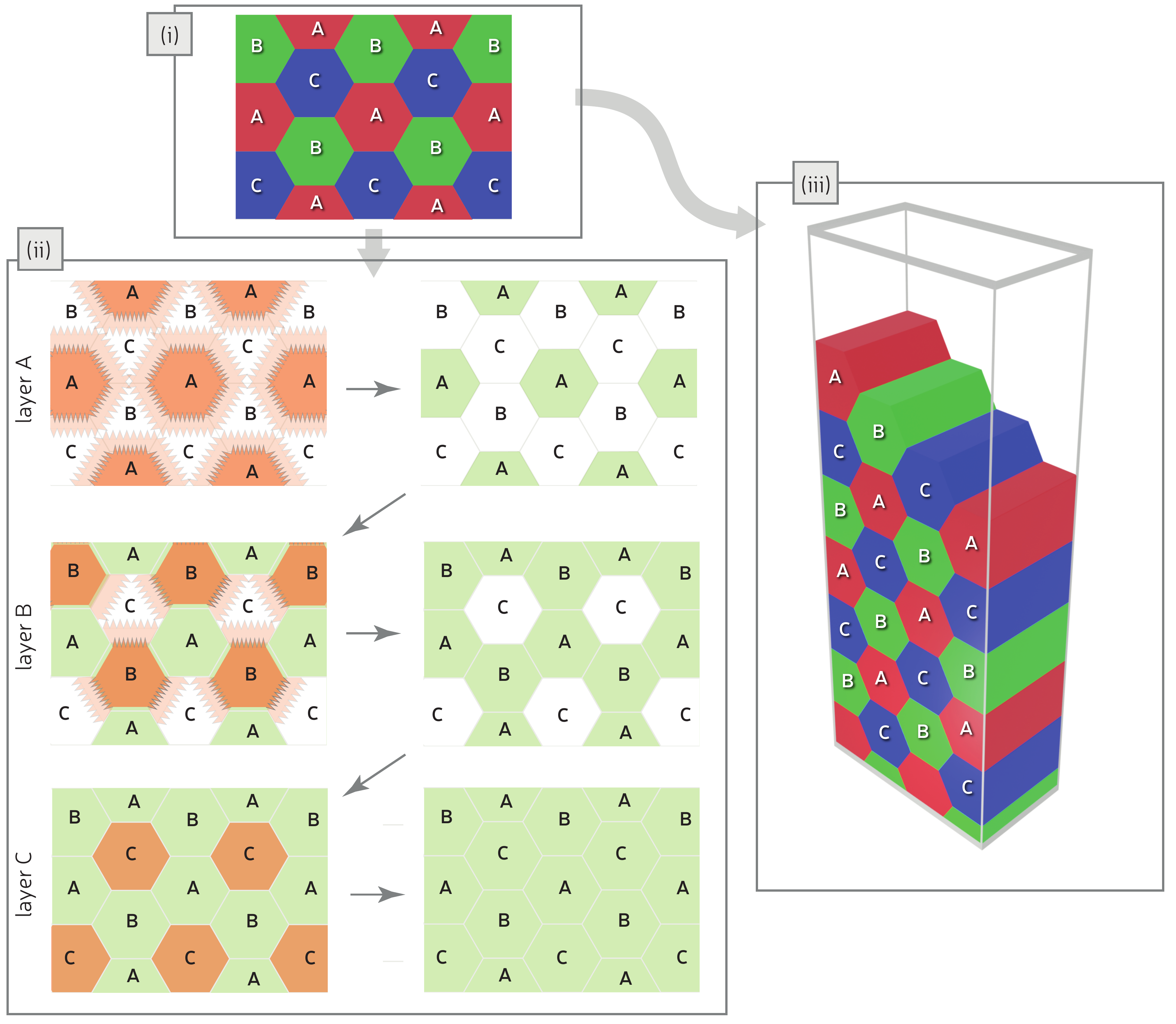}
    \caption{Parallel window decoding in both time and 1 spatial dimension and the relationship to colourability of tessellations. (\textit{i}) A 3-colour hexagonal tessellation of a 2D space, with each colour assigned a layer label A, B or C.  Note that hexagons of the same colour never touch.  (\textit{ii}) A protocol (in 2D) based on the hexagonal tiling.  The colours here match those used in \cref{fig:ParallelWindow}.  That is, dark orange indicates a commit region and light orange shows the buffer region.  Zig-zag boundaries represent rough boundaries.  Green indicates regions where all the defects have been resolved. (\textit{iii}) The hexagonal pattern of (\textit{i}) extruded into the 3rd dimension, so it is suitable for surface code decoding (e.g 2D+1 decoding problems).}
    \label{fig:3layersIn2D}
\end{figure*}

\section{Parallel window decoding in time and space}
\label{App:higherDim}

Our main argument has centered around how to perform parallel window decoding over windows defined by time intervals.  However, as motivated in the main text, we also may want to parallelize with respect to spatial directions.  This is required to support long range lattice surgery operations~\cite{horsman2012surface,fowler2018low,litinski2019game,chamberland2022circuit,chamberland2022universal}, and may also be desirable within single patches.  Here we outline how this works, with a guiding example given in \cref{fig:3layersIn2D}.

First, given some space (e.g. a decoding graph or hypergraph) we divide the space up into non-overlapping commit regions.  We regard each vertex in the decoding problem as having a space-time coordinate in $\mathbb{R}^D$ (with $D=2$ for the surface code).  Each edge in the decoding graph is assigned a space-time coordinate corresponding to the mid-point between the vertices it connects.  For edges connecting to the boundary, we can just equate the non-boundary vertex coordinate with the edge coordinate. Then for any space-time region, we can associate a set of vertices and edges residing within this region.  Assuming a topological code that has local stabilizers, then will always be a maximum distance $R$ between any pair of vertices connected by an edge.

Therefore, to find a valid ordering of layers, it suffices to solve a colouring problem.  That is, we define collections of commit regions and seek to assign them colours, such that (i) no two regions of the same colour are adjacent; (ii) length scales are set so that regions of the same colour are always separated by distance $R$. Given such a colouring, we can map colours to decoding layers, for example
$\mathrm{red} \rightarrow A$, $\mathrm{green} \rightarrow B$ and $\mathrm{blue} \rightarrow C$.  Any permutation of layers remains a valid choice.

We can regard commit regions $A$ and $B$ of \cref{fig:ParallelWindow} as representing a 2-colouring of a 2D space.  This is extended to 3D (and thereby the surface code decoding problem) by extruding into a 3rd dimension.  \cref{fig:3layersIn2D}-i shows a hexagonal 3-colouring of a 2D space, and \cref{fig:3layersIn2D}-iii shows the extruded 3D version of this tiling.  For a $D$ dimensional space there exist tilings that can be coloured using $D+1$ colours with each tile of bounded size, which for instance has been proved in the context of colour codes \cite{bombin2013introduction,KubicaBeverland15}.  In \cref{fig:3layersIn2D}-iii, we tile a $D=3$ space using only 3 colours, but the regions are unbounded size with respect to depth in the 3rd dimension.  If we desire constant size tiles, then a tiling of 3D space could be achieved using 4 colours.

Our examples show the minimum number of colours.  Given a limited number of processors $N_{\mathrm{par}}$, we may choose to use more colours so that for each colour there are no more than $N_{\mathrm{par}}$ regions.

Next, we consider the buffer regions required to provide confidence in the corrections in the commit regions.  In \cref{fig:ParallelWindow}, the buffer windows are placed above and below the commit region of layer A.  In higher dimensions, the buffer regions must include all possible error locations (edges) within a distance $w$ of the commit region.  However, previously committed regions must not be included in the construction of buffers.  Additionally,  we do not want artificial defects pushed into a previously resolved region. Therefore, where a window meets a previously committed region the boundary must be set to smooth (no artificial defects allowed).

For example, \cref{fig:3layersIn2D}-ii shows buffer regions and boundaries for a hexagonal tiling.  In layer $A$, the buffer region extends in every direction from the commit region.  All the boundaries in $A$ are rough.   In layer $B$, the buffer extends in all directions except those already resolved in layer $A$.  Furthermore, the layer $B$ window boundaries are set rough except where they meet the resolved layer $A$ commit regions (where they are instead smooth, as illustrated).  The final layer $C$ will only have smooth boundaries and no buffer regions.

\section{Circuit-level variants noise and non-matching inner decoders}
\label{App:nonmatching}

Here we outline how our sliding and parallel window methods generalise to circuit-level noise and non-matching decoding problems.  In circuit-level noise, we may have a decoding problem that possibly includes so-called hook errors, which can be represented by additional edges that are neither solely horizontal or vertical, but instead diagonal in an otherwise cubic graph.  

Non-matching decoding problems arise when we also include the possibility of hyperedges.  That is, given an error $E$, the associated hyperedge is a list of all the defects it triggers should the error occur. If this list of defects contains more than 2 elements, we say it is a hyperedge.  To extend our method, we simply partition all the hyperedges into sets that we call commit regions.  Two commit regions can be of the same colour (and therefore part of the same layer) provided that there is no vertex/defect contained in hyperedges from both sets.  

This partitioning can be performed either by defining a distance metric derived from the hypergraph, or by defining space-time regions and with each hyperedge having a space-time co-ordinate based on the mid-point of its associated vertices. Note that for non-topological codes the decoding hypergraph may not be localized in Euclidean space, though repeated syndrome extraction means that there will be a time axis such that hyperedges contain vertices that are contained within a constant range on the time axis.

For buffer regions, we follow the same recipe as in the matching case.  The difference between rough and smooth boundaries needs additional care. Wherever we have a rough boundary (extremal hyperedges in a buffer region that are not adjacent to any previously corrected/committed regions), we need to allow for the possibility of creating artificial defects.  This can be achieved by connecting every hyperedge on a rough boundary to the boundary vertex.

\section{Numerical validation of decoder performance}
\label{app:numerical_validation}

In the main text, we presented numerical results for parallel window decoding using a MWPM inner decoder.  Here we present and discuss some additional numerical results: the performance of sliding window decoders with a MWPM inner decoder; and parallel window decoding with a UF inner decoder.

\begin{figure*}[t]
    \centering
    \includegraphics[width=450pt]{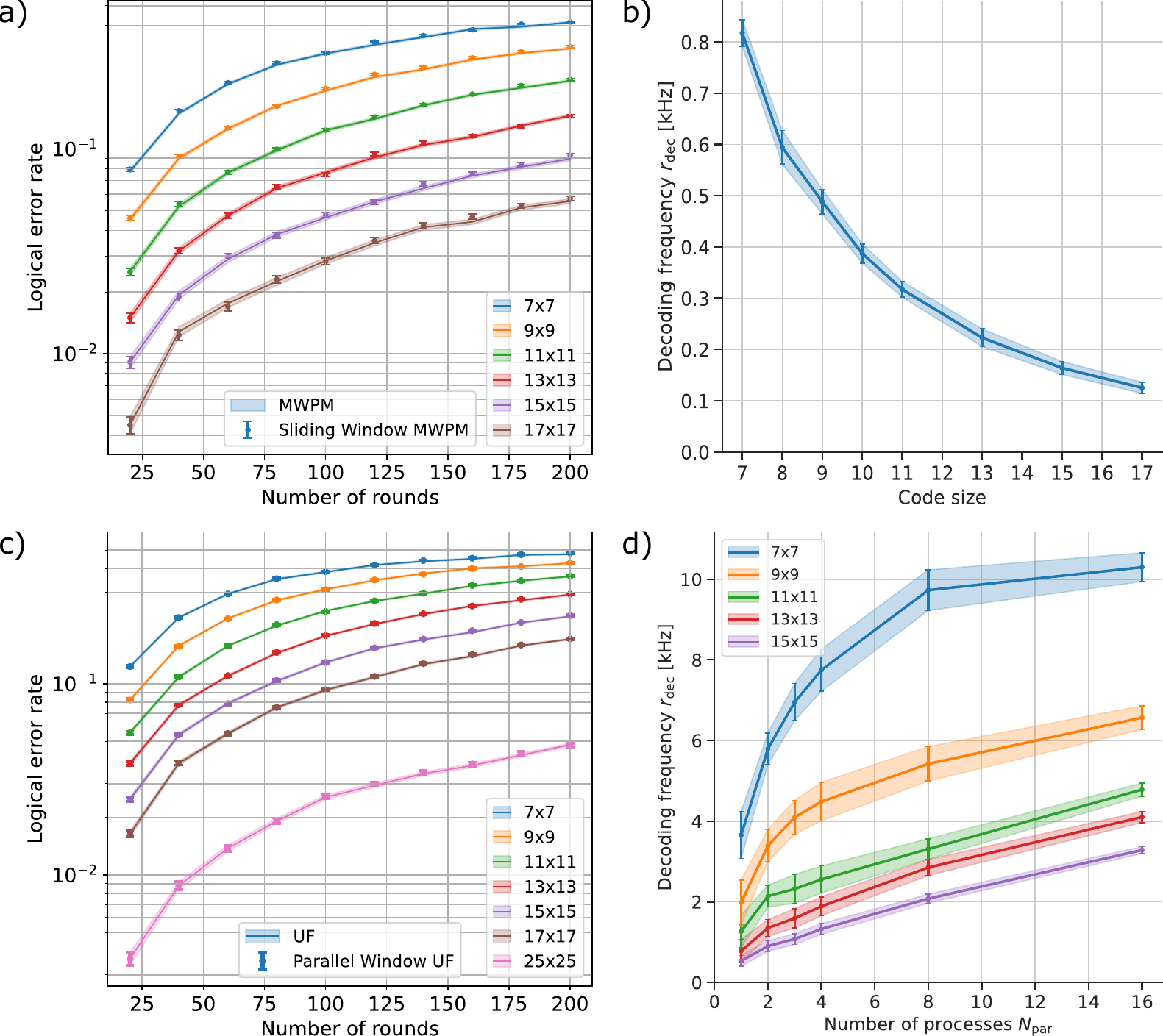}
    \caption{Logical error rate and decoding frequency on rotated planar code using sliding window MWPM decoder, and parallel window decoder with union-find under phenomenological Pauli noise with 2\% physical error rate.
        (a) Logical error rates as a function of the number of rounds of syndrome extraction for different code sizes for the global MWPM (lines), and using the sliding window MWPM decoder (points).
        (b) The decoding frequency as a function of the code size $d$ for square rotated planar codes using a sliding window MWPM decoder.
        (c) Logical error rates as a function of the number of rounds for global UF (lines) and using the parallel window algorithm with UF inner decoder (points).
        (d) The decoding frequency as a function of the number of decoding processes for the parallel window UF algorithm.
        Where the error bars are not visible, they are smaller than the marker size. Here we plot the decoding frequency $r_{\mathrm{dec}}$, therefore the rate of syndrome processing is $r_{\mathrm{proc}} =  r_{\mathrm{dec}} (d^2-1) $}
    \label{fig:SlidingWindowUF}
\end{figure*}

In \cref{fig:SlidingWindowUF}a we confirm that the sliding window decoding has a negligible drop in logical fidelity for $n_{\mathrm{W}} = 2d$,  $n_{\mathrm{com}} = d$ when compared to the global MWPM decoder.
Furthermore, in \cref{fig:SlidingWindowUF}b we measure the decoding frequency as a function of code size for square rotated planar codes. As the code size grows, the decoding frequency is expected to reduce as $O(1/\mathrm{poly}(d))$ for both MWPM and UF which is consistent with our data. Therefore, using sliding window decoding combined with any of the leading inner decoding algorithms, there will always be a code distance for which $\tau_{\mathrm{W}} > n_{\mathrm{com}} \tau_{\mathrm{rd}}$. This sets a limit on the distance up to which error correction codes can scale using sliding window decoding.

Next, we discuss parallel window decoding when UF replaces MWPM as the inner decoder. As with MWPM, we see no significant increase of the logical error rate when using parallel window decoding (\cref{fig:SlidingWindowUF}c), and a roughly linear increase with the number of processes $N_{\mathrm{par}}$ for large codes. However, in the case of smaller codes the decoding problem is relatively easy and we see diminishing returns with increased parallelism as the parallelization overheads in Python start being comparable with the decoding time of individual windows.

Sending data to a worker process, starting the decoding of a window and receiving the resulting data takes a finite amount of time $\tau_0$. Therefore, if $N_\mathrm{par}\tau_0 > \tau_\mathrm{W}$ all parallel processes will never be fully utilized and the processing will be bottle-necked by these overheads. However, in a hardware decoder, we expect $\tau_0$ to be below 10ns using modern hardware and syndrome compression techniques \cite{das2020scalable}, allowing us to scale to over $100$ processes. As separate processes do not need to share data, further parallelization of data communication is possible, allowing for even higher bandwidths.

\section{Decoding pipeline}
\label{app:pipelining}

\begin{figure*}[t]
    \centering
    \includegraphics[width=500pt]{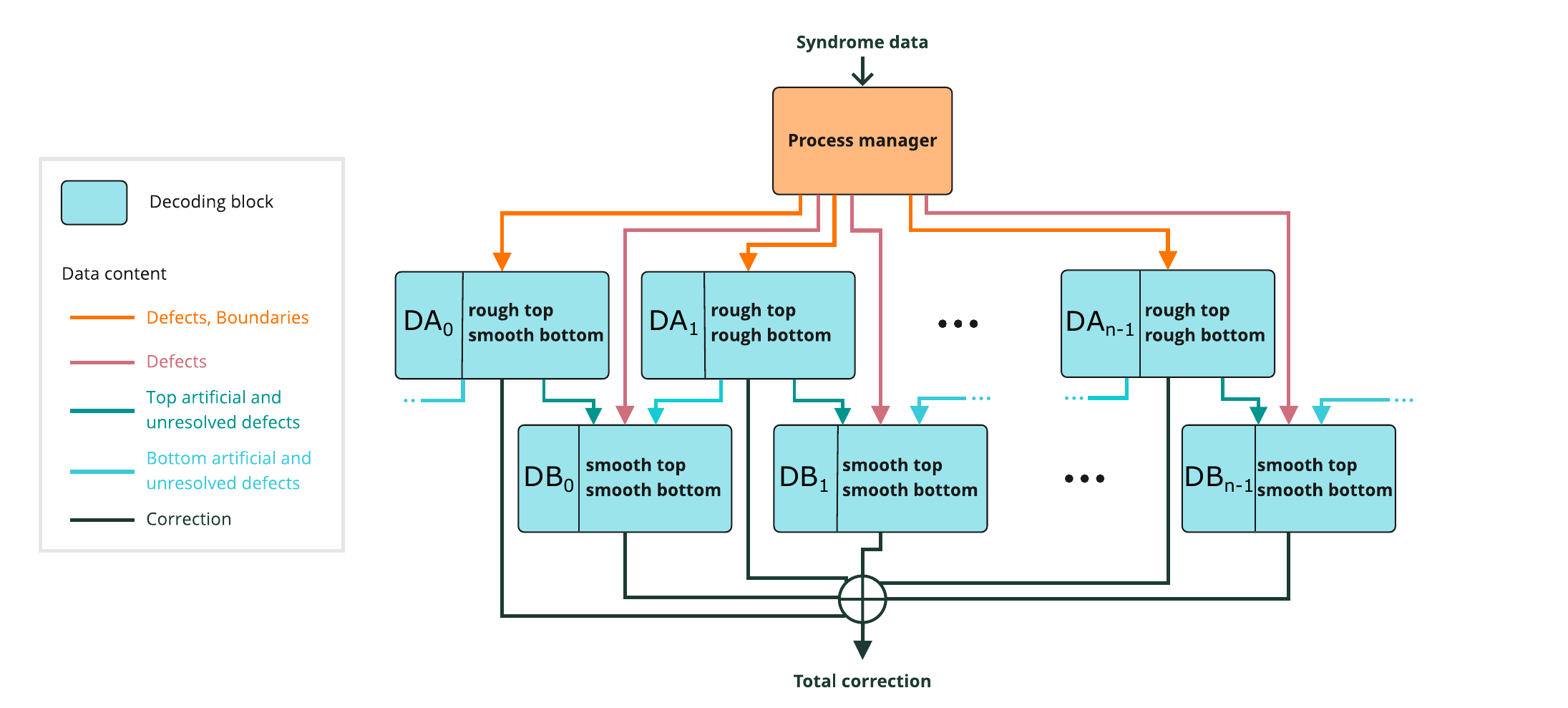}
    \caption{Parallel window decoding pipeline. The content of the data lines is colour-coded and described in the legend on the left. Each decoding block implements a matching algorithm on $3d$ rounds with specified time boundaries. The process manager can control the time boundaries of $DA_i$ blocks to match the global initial and final rounds. The blocks are connected cyclically as the line going from $DA_0$ to the left is connected to the line to $DB_{n-1}$ coming from the right.}
    \label{fig:Pipelining}
\end{figure*}

In \cref{fig:Pipelining}, we sketch the data-flow of the parallel window decoder with $2n$ processes that could be implemented in hardware. As the stream of syndrome data is acquired, it is given to the process manager that is in charge of passing the data to the appropriate decoding block.
Each decoding block resolves the $3d$ rounds of defects given to it using a matching decoder of choice, and a given specification of rough time boundaries. Recall that the decode windows are labelled $A_i$ and $B_i$ respectively, where $i$ is an integer. 
%
%
In the $k$-th step, $DA_i$ ($DB_i$) blocks decode windows $A_{kn+i}$ ($B_{kn+i}$) which have rough (smooth) time boundaries (see \cref{fig:ParallelWindow}). The exception are the first and the last blocks of computation whose boundaries depend on the initialized state and the basis of measurement.

When the first $3d$ rounds have been collected, these are sent to block $DA_0$ for decoding, together with the bottom boundary-type information.
The next $d$ rounds are given to $DB_0$ block which has to wait for $DA_0$ and $DA_1$ to finish before starting, followed by $3d$ rounds for block $DA_1$ and continuing until all blocks are running.
Once $DA_i$ finishes decoding, it sends the artificial defects and unresolved syndromes from the bottom $d$ rounds to $DB_{i-1}$, and from the top $d$ rounds to $DB_{i}$. The indices are cyclic with period $n$, meaning that $DB_{-1}=DB_{n-1}$, and $DA_{n-1}$ block is followed by $DA_0$.
When the data from $DA_i$ and $DA_{i+1}$ has been received, the $DB_i$ block can start decoding.
The committed corrections from all blocks are added together, continuously updating the total correction.

\bibliography{Refs}

\begin{thebibliography}{45}%
\makeatletter
\providecommand \@ifxundefined [1]{%
 \@ifx{#1\undefined}
}%
\providecommand \@ifnum [1]{%
 \ifnum #1\expandafter \@firstoftwo
 \else \expandafter \@secondoftwo
 \fi
}%
\providecommand \@ifx [1]{%
 \ifx #1\expandafter \@firstoftwo
 \else \expandafter \@secondoftwo
 \fi
}%
\providecommand \natexlab [1]{#1}%
\providecommand \enquote  [1]{``#1''}%
\providecommand \bibnamefont  [1]{#1}%
\providecommand \bibfnamefont [1]{#1}%
\providecommand \citenamefont [1]{#1}%
\providecommand \href@noop [0]{\@secondoftwo}%
\providecommand \href [0]{\begingroup \@sanitize@url \@href}%
\providecommand \@href[1]{\@@startlink{#1}\@@href}%
\providecommand \@@href[1]{\endgroup#1\@@endlink}%
\providecommand \@sanitize@url [0]{\catcode `\\12\catcode `\$12\catcode
  `\&12\catcode `\#12\catcode `\^12\catcode `\_12\catcode `\%12\relax}%
\providecommand \@@startlink[1]{}%
\providecommand \@@endlink[0]{}%
\providecommand \url  [0]{\begingroup\@sanitize@url \@url }%
\providecommand \@url [1]{\endgroup\@href {#1}{\urlprefix }}%
\providecommand \urlprefix  [0]{URL }%
\providecommand \Eprint [0]{\href }%
\providecommand \doibase [0]{https://doi.org/}%
\providecommand \selectlanguage [0]{\@gobble}%
\providecommand \bibinfo  [0]{\@secondoftwo}%
\providecommand \bibfield  [0]{\@secondoftwo}%
\providecommand \translation [1]{[#1]}%
\providecommand \BibitemOpen [0]{}%
\providecommand \bibitemStop [0]{}%
\providecommand \bibitemNoStop [0]{.\EOS\space}%
\providecommand \EOS [0]{\spacefactor3000\relax}%
\providecommand \BibitemShut  [1]{\csname bibitem#1\endcsname}%
\let\auto@bib@innerbib\@empty
\bibitem [{\citenamefont {Terhal}(2015)}]{terhal2015quantum}%
  \BibitemOpen
  \bibfield  {author} {\bibinfo {author} {\bibfnamefont {B.~M.}\ \bibnamefont
  {Terhal}},\ }\bibfield  {title} {\bibinfo {title} {Quantum error correction
  for quantum memories},\ }\href@noop {} {\bibfield  {journal} {\bibinfo
  {journal} {Reviews of Modern Physics}\ }\textbf {\bibinfo {volume} {87}},\
  \bibinfo {pages} {307} (\bibinfo {year} {2015})}\BibitemShut {NoStop}%
\bibitem [{\citenamefont {Bravyi}\ and\ \citenamefont
  {Kitaev}(2005)}]{bravyi2005universal}%
  \BibitemOpen
  \bibfield  {author} {\bibinfo {author} {\bibfnamefont {S.}~\bibnamefont
  {Bravyi}}\ and\ \bibinfo {author} {\bibfnamefont {A.}~\bibnamefont
  {Kitaev}},\ }\bibfield  {title} {\bibinfo {title} {Universal quantum
  computation with ideal clifford gates and noisy ancillas},\ }\href@noop {}
  {\bibfield  {journal} {\bibinfo  {journal} {Physical Review A}\ }\textbf
  {\bibinfo {volume} {71}},\ \bibinfo {pages} {022316} (\bibinfo {year}
  {2005})}\BibitemShut {NoStop}%
\bibitem [{\citenamefont {Litinski}(2019)}]{litinski2019game}%
  \BibitemOpen
  \bibfield  {author} {\bibinfo {author} {\bibfnamefont {D.}~\bibnamefont
  {Litinski}},\ }\bibfield  {title} {\bibinfo {title} {A game of surface codes:
  Large-scale quantum computing with lattice surgery},\ }\href@noop {}
  {\bibfield  {journal} {\bibinfo  {journal} {Quantum}\ }\textbf {\bibinfo
  {volume} {3}},\ \bibinfo {pages} {128} (\bibinfo {year} {2019})}\BibitemShut
  {NoStop}%
\bibitem [{\citenamefont {DiVincenzo}\ and\ \citenamefont
  {Aliferis}(2007)}]{divincenzo2007slow}%
  \BibitemOpen
  \bibfield  {author} {\bibinfo {author} {\bibfnamefont {D.~P.}\ \bibnamefont
  {DiVincenzo}}\ and\ \bibinfo {author} {\bibfnamefont {P.}~\bibnamefont
  {Aliferis}},\ }\bibfield  {title} {\bibinfo {title} {Effective fault-tolerant
  quantum computation with slow measurements},\ }\href
  {https://doi.org/10.1103/PhysRevLett.98.020501} {\bibfield  {journal}
  {\bibinfo  {journal} {Phys. Rev. Lett.}\ }\textbf {\bibinfo {volume} {98}},\
  \bibinfo {pages} {020501} (\bibinfo {year} {2007})}\BibitemShut {NoStop}%
\bibitem [{\citenamefont {Chamberland}\ \emph {et~al.}(2018)\citenamefont
  {Chamberland}, \citenamefont {Iyer},\ and\ \citenamefont
  {Poulin}}]{Chamberland2018faulttolerant}%
  \BibitemOpen
  \bibfield  {author} {\bibinfo {author} {\bibfnamefont {C.}~\bibnamefont
  {Chamberland}}, \bibinfo {author} {\bibfnamefont {P.}~\bibnamefont {Iyer}},\
  and\ \bibinfo {author} {\bibfnamefont {D.}~\bibnamefont {Poulin}},\
  }\bibfield  {title} {\bibinfo {title} {Fault-tolerant quantum computing in
  the {P}auli or {C}lifford frame with slow error diagnostics},\ }\href
  {https://doi.org/10.22331/q-2018-01-04-43} {\bibfield  {journal} {\bibinfo
  {journal} {{Quantum}}\ }\textbf {\bibinfo {volume} {2}},\ \bibinfo {pages}
  {43} (\bibinfo {year} {2018})}\BibitemShut {NoStop}%
\bibitem [{\citenamefont {Holmes}\ \emph {et~al.}(2020)\citenamefont {Holmes},
  \citenamefont {Jokar}, \citenamefont {Pasandi}, \citenamefont {Ding},
  \citenamefont {Pedram},\ and\ \citenamefont {Chong}}]{holmes2020nisq}%
  \BibitemOpen
  \bibfield  {author} {\bibinfo {author} {\bibfnamefont {A.}~\bibnamefont
  {Holmes}}, \bibinfo {author} {\bibfnamefont {M.~R.}\ \bibnamefont {Jokar}},
  \bibinfo {author} {\bibfnamefont {G.}~\bibnamefont {Pasandi}}, \bibinfo
  {author} {\bibfnamefont {Y.}~\bibnamefont {Ding}}, \bibinfo {author}
  {\bibfnamefont {M.}~\bibnamefont {Pedram}},\ and\ \bibinfo {author}
  {\bibfnamefont {F.~T.}\ \bibnamefont {Chong}},\ }\bibfield  {title} {\bibinfo
  {title} {Nisq+: Boosting quantum computing power by approximating quantum
  error correction},\ }in\ \href@noop {} {\emph {\bibinfo {booktitle} {2020
  ACM/IEEE 47th Annual International Symposium on Computer Architecture
  (ISCA)}}}\ (\bibinfo {organization} {IEEE},\ \bibinfo {year} {2020})\ pp.\
  \bibinfo {pages} {556--569}\BibitemShut {NoStop}%
\bibitem [{\citenamefont {Chamberland}\ \emph
  {et~al.}(2022{\natexlab{a}})\citenamefont {Chamberland}, \citenamefont
  {Goncalves}, \citenamefont {Sivarajah}, \citenamefont {Peterson},\ and\
  \citenamefont {Grimberg}}]{chamberland2022techniques}%
  \BibitemOpen
  \bibfield  {author} {\bibinfo {author} {\bibfnamefont {C.}~\bibnamefont
  {Chamberland}}, \bibinfo {author} {\bibfnamefont {L.}~\bibnamefont
  {Goncalves}}, \bibinfo {author} {\bibfnamefont {P.}~\bibnamefont
  {Sivarajah}}, \bibinfo {author} {\bibfnamefont {E.}~\bibnamefont
  {Peterson}},\ and\ \bibinfo {author} {\bibfnamefont {S.}~\bibnamefont
  {Grimberg}},\ }\bibfield  {title} {\bibinfo {title} {Techniques for combining
  fast local decoders with global decoders under circuit-level noise},\
  }\href@noop {} {\bibfield  {journal} {\bibinfo  {journal} {arXiv preprint
  arXiv:2208.01178}\ } (\bibinfo {year} {2022}{\natexlab{a}})}\BibitemShut
  {NoStop}%
\bibitem [{\citenamefont {Dennis}\ \emph {et~al.}(2002)\citenamefont {Dennis},
  \citenamefont {Kitaev}, \citenamefont {Landahl},\ and\ \citenamefont
  {Preskill}}]{dennis2002topological}%
  \BibitemOpen
  \bibfield  {author} {\bibinfo {author} {\bibfnamefont {E.}~\bibnamefont
  {Dennis}}, \bibinfo {author} {\bibfnamefont {A.}~\bibnamefont {Kitaev}},
  \bibinfo {author} {\bibfnamefont {A.}~\bibnamefont {Landahl}},\ and\ \bibinfo
  {author} {\bibfnamefont {J.}~\bibnamefont {Preskill}},\ }\bibfield  {title}
  {\bibinfo {title} {Topological quantum memory},\ }\href@noop {} {\bibfield
  {journal} {\bibinfo  {journal} {Journal of Mathematical Physics}\ }\textbf
  {\bibinfo {volume} {43}},\ \bibinfo {pages} {4452} (\bibinfo {year}
  {2002})}\BibitemShut {NoStop}%
\bibitem [{\citenamefont {Fowler}\ \emph {et~al.}(2009)\citenamefont {Fowler},
  \citenamefont {Stephens},\ and\ \citenamefont
  {Groszkowski}}]{fowler2009high}%
  \BibitemOpen
  \bibfield  {author} {\bibinfo {author} {\bibfnamefont {A.~G.}\ \bibnamefont
  {Fowler}}, \bibinfo {author} {\bibfnamefont {A.~M.}\ \bibnamefont
  {Stephens}},\ and\ \bibinfo {author} {\bibfnamefont {P.}~\bibnamefont
  {Groszkowski}},\ }\bibfield  {title} {\bibinfo {title} {High-threshold
  universal quantum computation on the surface code},\ }\href@noop {}
  {\bibfield  {journal} {\bibinfo  {journal} {Physical Review A}\ }\textbf
  {\bibinfo {volume} {80}},\ \bibinfo {pages} {052312} (\bibinfo {year}
  {2009})}\BibitemShut {NoStop}%
\bibitem [{\citenamefont {Higgott}(2021)}]{higgott2021pymatching}%
  \BibitemOpen
  \bibfield  {author} {\bibinfo {author} {\bibfnamefont {O.}~\bibnamefont
  {Higgott}},\ }\bibfield  {title} {\bibinfo {title} {{PyMatching}: A python
  package for decoding quantum codes with minimum-weight perfect matching},\
  }\href@noop {} {\bibfield  {journal} {\bibinfo  {journal} {arXiv preprint
  arXiv:2105.13082}\ } (\bibinfo {year} {2021})}\BibitemShut {NoStop}%
\bibitem [{\citenamefont {Delfosse}\ and\ \citenamefont
  {Nickerson}(2021)}]{delfosse2021almost}%
  \BibitemOpen
  \bibfield  {author} {\bibinfo {author} {\bibfnamefont {N.}~\bibnamefont
  {Delfosse}}\ and\ \bibinfo {author} {\bibfnamefont {N.~H.}\ \bibnamefont
  {Nickerson}},\ }\bibfield  {title} {\bibinfo {title} {Almost-linear time
  decoding algorithm for topological codes},\ }\href@noop {} {\bibfield
  {journal} {\bibinfo  {journal} {Quantum}\ }\textbf {\bibinfo {volume} {5}},\
  \bibinfo {pages} {595} (\bibinfo {year} {2021})}\BibitemShut {NoStop}%
\bibitem [{\citenamefont {Das}\ \emph {et~al.}(2020)\citenamefont {Das},
  \citenamefont {Pattison}, \citenamefont {Manne}, \citenamefont {Carmean},
  \citenamefont {Svore}, \citenamefont {Qureshi},\ and\ \citenamefont
  {Delfosse}}]{das2020scalable}%
  \BibitemOpen
  \bibfield  {author} {\bibinfo {author} {\bibfnamefont {P.}~\bibnamefont
  {Das}}, \bibinfo {author} {\bibfnamefont {C.~A.}\ \bibnamefont {Pattison}},
  \bibinfo {author} {\bibfnamefont {S.}~\bibnamefont {Manne}}, \bibinfo
  {author} {\bibfnamefont {D.}~\bibnamefont {Carmean}}, \bibinfo {author}
  {\bibfnamefont {K.}~\bibnamefont {Svore}}, \bibinfo {author} {\bibfnamefont
  {M.}~\bibnamefont {Qureshi}},\ and\ \bibinfo {author} {\bibfnamefont
  {N.}~\bibnamefont {Delfosse}},\ }\bibfield  {title} {\bibinfo {title} {A
  scalable decoder micro-architecture for fault-tolerant quantum computing},\
  }\href@noop {} {\bibfield  {journal} {\bibinfo  {journal} {arXiv preprint
  arXiv:2001.06598}\ } (\bibinfo {year} {2020})}\BibitemShut {NoStop}%
\bibitem [{\citenamefont {Huang}\ and\ \citenamefont
  {Brown}(2021)}]{huang2021between}%
  \BibitemOpen
  \bibfield  {author} {\bibinfo {author} {\bibfnamefont {S.}~\bibnamefont
  {Huang}}\ and\ \bibinfo {author} {\bibfnamefont {K.~R.}\ \bibnamefont
  {Brown}},\ }\bibfield  {title} {\bibinfo {title} {Between shor and steane: A
  unifying construction for measuring error syndromes},\ }\href@noop {}
  {\bibfield  {journal} {\bibinfo  {journal} {Physical Review Letters}\
  }\textbf {\bibinfo {volume} {127}},\ \bibinfo {pages} {090505} (\bibinfo
  {year} {2021})}\BibitemShut {NoStop}%
\bibitem [{\citenamefont {Iyengar}\ \emph {et~al.}(2012)\citenamefont
  {Iyengar}, \citenamefont {Papaleo}, \citenamefont {Siegel}, \citenamefont
  {Wolf}, \citenamefont {Vanelli-Coralli},\ and\ \citenamefont
  {Corazza}}]{Iyengar2012sliding}%
  \BibitemOpen
  \bibfield  {author} {\bibinfo {author} {\bibfnamefont {A.~R.}\ \bibnamefont
  {Iyengar}}, \bibinfo {author} {\bibfnamefont {M.}~\bibnamefont {Papaleo}},
  \bibinfo {author} {\bibfnamefont {P.~H.}\ \bibnamefont {Siegel}}, \bibinfo
  {author} {\bibfnamefont {J.~K.}\ \bibnamefont {Wolf}}, \bibinfo {author}
  {\bibfnamefont {A.}~\bibnamefont {Vanelli-Coralli}},\ and\ \bibinfo {author}
  {\bibfnamefont {G.~E.}\ \bibnamefont {Corazza}},\ }\bibfield  {title}
  {\bibinfo {title} {Windowed decoding of protograph-based {LDPC} convolutional
  codes over erasure channels},\ }\href
  {https://doi.org/10.1109/TIT.2011.2177439} {\bibfield  {journal} {\bibinfo
  {journal} {IEEE Transactions on Information Theory}\ }\textbf {\bibinfo
  {volume} {58}},\ \bibinfo {pages} {2303} (\bibinfo {year}
  {2012})}\BibitemShut {NoStop}%
\bibitem [{\citenamefont {Tan}\ \emph {et~al.}(2022)\citenamefont {Tan},
  \citenamefont {Zhang}, \citenamefont {Chao}, \citenamefont {Shi},\ and\
  \citenamefont {Chen}}]{tan2022scalable}%
  \BibitemOpen
  \bibfield  {author} {\bibinfo {author} {\bibfnamefont {X.}~\bibnamefont
  {Tan}}, \bibinfo {author} {\bibfnamefont {F.}~\bibnamefont {Zhang}}, \bibinfo
  {author} {\bibfnamefont {R.}~\bibnamefont {Chao}}, \bibinfo {author}
  {\bibfnamefont {Y.}~\bibnamefont {Shi}},\ and\ \bibinfo {author}
  {\bibfnamefont {J.}~\bibnamefont {Chen}},\ }\bibfield  {title} {\bibinfo
  {title} {Scalable surface code decoders with parallelization in time},\
  }\href@noop {} {\bibfield  {journal} {\bibinfo  {journal} {arXiv preprint
  arXiv:2209.09219}\ } (\bibinfo {year} {2022})}\BibitemShut {NoStop}%
\bibitem [{\citenamefont {Fowler}(2012)}]{austinfowlertimeoptimalpaper}%
  \BibitemOpen
  \bibfield  {author} {\bibinfo {author} {\bibfnamefont {A.~G.}\ \bibnamefont
  {Fowler}},\ }\bibfield  {title} {\bibinfo {title} {Time-optimal quantum
  computation},\ }\href@noop {} {\bibfield  {journal} {\bibinfo  {journal}
  {arXiv preprint arXiv:1210.4626}\ } (\bibinfo {year} {2012})}\BibitemShut
  {NoStop}%
\bibitem [{\citenamefont {Gidney}\ and\ \citenamefont
  {Fowler}(2019)}]{gidney2019flexible}%
  \BibitemOpen
  \bibfield  {author} {\bibinfo {author} {\bibfnamefont {C.}~\bibnamefont
  {Gidney}}\ and\ \bibinfo {author} {\bibfnamefont {A.~G.}\ \bibnamefont
  {Fowler}},\ }\bibfield  {title} {\bibinfo {title} {Flexible layout of surface
  code computations using autoccz states},\ }\href@noop {} {\bibfield
  {journal} {\bibinfo  {journal} {arXiv preprint arXiv:1905.08916}\ } (\bibinfo
  {year} {2019})}\BibitemShut {NoStop}%
\bibitem [{\citenamefont {Berry}\ \emph {et~al.}(2019)\citenamefont {Berry},
  \citenamefont {Gidney}, \citenamefont {Motta}, \citenamefont {McClean},\ and\
  \citenamefont {Babbush}}]{berry2019qubitization}%
  \BibitemOpen
  \bibfield  {author} {\bibinfo {author} {\bibfnamefont {D.~W.}\ \bibnamefont
  {Berry}}, \bibinfo {author} {\bibfnamefont {C.}~\bibnamefont {Gidney}},
  \bibinfo {author} {\bibfnamefont {M.}~\bibnamefont {Motta}}, \bibinfo
  {author} {\bibfnamefont {J.~R.}\ \bibnamefont {McClean}},\ and\ \bibinfo
  {author} {\bibfnamefont {R.}~\bibnamefont {Babbush}},\ }\bibfield  {title}
  {\bibinfo {title} {Qubitization of arbitrary basis quantum chemistry
  leveraging sparsity and low rank factorization},\ }\href@noop {} {\bibfield
  {journal} {\bibinfo  {journal} {Quantum}\ }\textbf {\bibinfo {volume} {3}},\
  \bibinfo {pages} {208} (\bibinfo {year} {2019})}\BibitemShut {NoStop}%
\bibitem [{\citenamefont {Kivlichan}\ \emph {et~al.}(2020)\citenamefont
  {Kivlichan}, \citenamefont {Gidney}, \citenamefont {Berry}, \citenamefont
  {Wiebe}, \citenamefont {McClean}, \citenamefont {Sun}, \citenamefont {Jiang},
  \citenamefont {Rubin}, \citenamefont {Fowler}, \citenamefont {Aspuru-Guzik}
  \emph {et~al.}}]{kivlichan2020improved}%
  \BibitemOpen
  \bibfield  {author} {\bibinfo {author} {\bibfnamefont {I.~D.}\ \bibnamefont
  {Kivlichan}}, \bibinfo {author} {\bibfnamefont {C.}~\bibnamefont {Gidney}},
  \bibinfo {author} {\bibfnamefont {D.~W.}\ \bibnamefont {Berry}}, \bibinfo
  {author} {\bibfnamefont {N.}~\bibnamefont {Wiebe}}, \bibinfo {author}
  {\bibfnamefont {J.}~\bibnamefont {McClean}}, \bibinfo {author} {\bibfnamefont
  {W.}~\bibnamefont {Sun}}, \bibinfo {author} {\bibfnamefont {Z.}~\bibnamefont
  {Jiang}}, \bibinfo {author} {\bibfnamefont {N.}~\bibnamefont {Rubin}},
  \bibinfo {author} {\bibfnamefont {A.}~\bibnamefont {Fowler}}, \bibinfo
  {author} {\bibfnamefont {A.}~\bibnamefont {Aspuru-Guzik}}, \emph {et~al.},\
  }\bibfield  {title} {\bibinfo {title} {Improved fault-tolerant quantum
  simulation of condensed-phase correlated electrons via trotterization},\
  }\href@noop {} {\bibfield  {journal} {\bibinfo  {journal} {Quantum}\ }\textbf
  {\bibinfo {volume} {4}},\ \bibinfo {pages} {296} (\bibinfo {year}
  {2020})}\BibitemShut {NoStop}%
\bibitem [{\citenamefont {Campbell}(2021)}]{campbell2021early}%
  \BibitemOpen
  \bibfield  {author} {\bibinfo {author} {\bibfnamefont {E.~T.}\ \bibnamefont
  {Campbell}},\ }\bibfield  {title} {\bibinfo {title} {Early fault-tolerant
  simulations of the hubbard model},\ }\href@noop {} {\bibfield  {journal}
  {\bibinfo  {journal} {Quantum Science and Technology}\ }\textbf {\bibinfo
  {volume} {7}},\ \bibinfo {pages} {015007} (\bibinfo {year}
  {2021})}\BibitemShut {NoStop}%
\bibitem [{\citenamefont {Chamberland}\ and\ \citenamefont
  {Campbell}(2022{\natexlab{a}})}]{chamberland2022universal}%
  \BibitemOpen
  \bibfield  {author} {\bibinfo {author} {\bibfnamefont {C.}~\bibnamefont
  {Chamberland}}\ and\ \bibinfo {author} {\bibfnamefont {E.~T.}\ \bibnamefont
  {Campbell}},\ }\bibfield  {title} {\bibinfo {title} {Universal quantum
  computing with twist-free and temporally encoded lattice surgery},\
  }\href@noop {} {\bibfield  {journal} {\bibinfo  {journal} {PRX Quantum}\
  }\textbf {\bibinfo {volume} {3}},\ \bibinfo {pages} {010331} (\bibinfo {year}
  {2022}{\natexlab{a}})}\BibitemShut {NoStop}%
\bibitem [{\citenamefont {Lee}\ \emph {et~al.}(2021)\citenamefont {Lee},
  \citenamefont {Berry}, \citenamefont {Gidney}, \citenamefont {Huggins},
  \citenamefont {McClean}, \citenamefont {Wiebe},\ and\ \citenamefont
  {Babbush}}]{lee2021even}%
  \BibitemOpen
  \bibfield  {author} {\bibinfo {author} {\bibfnamefont {J.}~\bibnamefont
  {Lee}}, \bibinfo {author} {\bibfnamefont {D.~W.}\ \bibnamefont {Berry}},
  \bibinfo {author} {\bibfnamefont {C.}~\bibnamefont {Gidney}}, \bibinfo
  {author} {\bibfnamefont {W.~J.}\ \bibnamefont {Huggins}}, \bibinfo {author}
  {\bibfnamefont {J.~R.}\ \bibnamefont {McClean}}, \bibinfo {author}
  {\bibfnamefont {N.}~\bibnamefont {Wiebe}},\ and\ \bibinfo {author}
  {\bibfnamefont {R.}~\bibnamefont {Babbush}},\ }\bibfield  {title} {\bibinfo
  {title} {Even more efficient quantum computations of chemistry through tensor
  hypercontraction},\ }\href@noop {} {\bibfield  {journal} {\bibinfo  {journal}
  {PRX Quantum}\ }\textbf {\bibinfo {volume} {2}},\ \bibinfo {pages} {030305}
  (\bibinfo {year} {2021})}\BibitemShut {NoStop}%
\bibitem [{\citenamefont {von Burg}\ \emph {et~al.}(2021)\citenamefont {von
  Burg}, \citenamefont {Low}, \citenamefont {H{\"a}ner}, \citenamefont
  {Steiger}, \citenamefont {Reiher}, \citenamefont {Roetteler},\ and\
  \citenamefont {Troyer}}]{von2021quantum}%
  \BibitemOpen
  \bibfield  {author} {\bibinfo {author} {\bibfnamefont {V.}~\bibnamefont {von
  Burg}}, \bibinfo {author} {\bibfnamefont {G.~H.}\ \bibnamefont {Low}},
  \bibinfo {author} {\bibfnamefont {T.}~\bibnamefont {H{\"a}ner}}, \bibinfo
  {author} {\bibfnamefont {D.~S.}\ \bibnamefont {Steiger}}, \bibinfo {author}
  {\bibfnamefont {M.}~\bibnamefont {Reiher}}, \bibinfo {author} {\bibfnamefont
  {M.}~\bibnamefont {Roetteler}},\ and\ \bibinfo {author} {\bibfnamefont
  {M.}~\bibnamefont {Troyer}},\ }\bibfield  {title} {\bibinfo {title} {Quantum
  computing enhanced computational catalysis},\ }\href@noop {} {\bibfield
  {journal} {\bibinfo  {journal} {Physical Review Research}\ }\textbf {\bibinfo
  {volume} {3}},\ \bibinfo {pages} {033055} (\bibinfo {year}
  {2021})}\BibitemShut {NoStop}%
\bibitem [{\citenamefont {Blunt}\ \emph {et~al.}(2022)\citenamefont {Blunt},
  \citenamefont {Camps}, \citenamefont {Crawford}, \citenamefont {Izs{\'a}k},
  \citenamefont {Leontica}, \citenamefont {Mirani}, \citenamefont {Moylett},
  \citenamefont {Scivier}, \citenamefont {S{\"u}nderhauf}, \citenamefont
  {Schopf} \emph {et~al.}}]{blunt2022perspective}%
  \BibitemOpen
  \bibfield  {author} {\bibinfo {author} {\bibfnamefont {N.~S.}\ \bibnamefont
  {Blunt}}, \bibinfo {author} {\bibfnamefont {J.}~\bibnamefont {Camps}},
  \bibinfo {author} {\bibfnamefont {O.}~\bibnamefont {Crawford}}, \bibinfo
  {author} {\bibfnamefont {R.}~\bibnamefont {Izs{\'a}k}}, \bibinfo {author}
  {\bibfnamefont {S.}~\bibnamefont {Leontica}}, \bibinfo {author}
  {\bibfnamefont {A.}~\bibnamefont {Mirani}}, \bibinfo {author} {\bibfnamefont
  {A.~E.}\ \bibnamefont {Moylett}}, \bibinfo {author} {\bibfnamefont {S.~A.}\
  \bibnamefont {Scivier}}, \bibinfo {author} {\bibfnamefont {C.}~\bibnamefont
  {S{\"u}nderhauf}}, \bibinfo {author} {\bibfnamefont {P.}~\bibnamefont
  {Schopf}}, \emph {et~al.},\ }\bibfield  {title} {\bibinfo {title} {A
  perspective on the current state-of-the-art of quantum computing for drug
  discovery applications},\ }\href@noop {} {\bibfield  {journal} {\bibinfo
  {journal} {arXiv preprint arXiv:2206.00551}\ } (\bibinfo {year}
  {2022})}\BibitemShut {NoStop}%
\bibitem [{\citenamefont {Chamberland}\ \emph
  {et~al.}(2022{\natexlab{b}})\citenamefont {Chamberland}, \citenamefont {Noh},
  \citenamefont {Arrangoiz-Arriola}, \citenamefont {Campbell}, \citenamefont
  {Hann}, \citenamefont {Iverson}, \citenamefont {Putterman}, \citenamefont
  {Bohdanowicz}, \citenamefont {Flammia}, \citenamefont {Keller} \emph
  {et~al.}}]{chamberland2022building}%
  \BibitemOpen
  \bibfield  {author} {\bibinfo {author} {\bibfnamefont {C.}~\bibnamefont
  {Chamberland}}, \bibinfo {author} {\bibfnamefont {K.}~\bibnamefont {Noh}},
  \bibinfo {author} {\bibfnamefont {P.}~\bibnamefont {Arrangoiz-Arriola}},
  \bibinfo {author} {\bibfnamefont {E.~T.}\ \bibnamefont {Campbell}}, \bibinfo
  {author} {\bibfnamefont {C.~T.}\ \bibnamefont {Hann}}, \bibinfo {author}
  {\bibfnamefont {J.}~\bibnamefont {Iverson}}, \bibinfo {author} {\bibfnamefont
  {H.}~\bibnamefont {Putterman}}, \bibinfo {author} {\bibfnamefont {T.~C.}\
  \bibnamefont {Bohdanowicz}}, \bibinfo {author} {\bibfnamefont {S.~T.}\
  \bibnamefont {Flammia}}, \bibinfo {author} {\bibfnamefont {A.}~\bibnamefont
  {Keller}}, \emph {et~al.},\ }\bibfield  {title} {\bibinfo {title} {Building a
  fault-tolerant quantum computer using concatenated cat codes},\ }\href@noop
  {} {\bibfield  {journal} {\bibinfo  {journal} {PRX Quantum}\ }\textbf
  {\bibinfo {volume} {3}},\ \bibinfo {pages} {010329} (\bibinfo {year}
  {2022}{\natexlab{b}})}\BibitemShut {NoStop}%
\bibitem [{\citenamefont {Darmawan}\ and\ \citenamefont
  {Poulin}(2017)}]{darmawan2017tensor}%
  \BibitemOpen
  \bibfield  {author} {\bibinfo {author} {\bibfnamefont {A.~S.}\ \bibnamefont
  {Darmawan}}\ and\ \bibinfo {author} {\bibfnamefont {D.}~\bibnamefont
  {Poulin}},\ }\bibfield  {title} {\bibinfo {title} {Tensor-network simulations
  of the surface code under realistic noise},\ }\href@noop {} {\bibfield
  {journal} {\bibinfo  {journal} {Physical review letters}\ }\textbf {\bibinfo
  {volume} {119}},\ \bibinfo {pages} {040502} (\bibinfo {year}
  {2017})}\BibitemShut {NoStop}%
\bibitem [{\citenamefont {Panteleev}\ and\ \citenamefont
  {Kalachev}(2021)}]{panteleev2021degenerate}%
  \BibitemOpen
  \bibfield  {author} {\bibinfo {author} {\bibfnamefont {P.}~\bibnamefont
  {Panteleev}}\ and\ \bibinfo {author} {\bibfnamefont {G.}~\bibnamefont
  {Kalachev}},\ }\bibfield  {title} {\bibinfo {title} {Degenerate quantum
  {LDPC} codes with good finite length performance},\ }\href@noop {} {\bibfield
   {journal} {\bibinfo  {journal} {Quantum}\ }\textbf {\bibinfo {volume} {5}},\
  \bibinfo {pages} {585} (\bibinfo {year} {2021})}\BibitemShut {NoStop}%
\bibitem [{\citenamefont {Roffe}\ \emph {et~al.}(2020)\citenamefont {Roffe},
  \citenamefont {White}, \citenamefont {Burton},\ and\ \citenamefont
  {Campbell}}]{roffe2020decoding}%
  \BibitemOpen
  \bibfield  {author} {\bibinfo {author} {\bibfnamefont {J.}~\bibnamefont
  {Roffe}}, \bibinfo {author} {\bibfnamefont {D.~R.}\ \bibnamefont {White}},
  \bibinfo {author} {\bibfnamefont {S.}~\bibnamefont {Burton}},\ and\ \bibinfo
  {author} {\bibfnamefont {E.}~\bibnamefont {Campbell}},\ }\bibfield  {title}
  {\bibinfo {title} {Decoding across the quantum low-density parity-check code
  landscape},\ }\href@noop {} {\bibfield  {journal} {\bibinfo  {journal}
  {Physical Review Research}\ }\textbf {\bibinfo {volume} {2}},\ \bibinfo
  {pages} {043423} (\bibinfo {year} {2020})}\BibitemShut {NoStop}%
\bibitem [{\citenamefont {Higgott}\ \emph {et~al.}(2022)\citenamefont
  {Higgott}, \citenamefont {Bohdanowicz}, \citenamefont {Kubica}, \citenamefont
  {Flammia},\ and\ \citenamefont {Campbell}}]{higgott2022fragile}%
  \BibitemOpen
  \bibfield  {author} {\bibinfo {author} {\bibfnamefont {O.}~\bibnamefont
  {Higgott}}, \bibinfo {author} {\bibfnamefont {T.~C.}\ \bibnamefont
  {Bohdanowicz}}, \bibinfo {author} {\bibfnamefont {A.}~\bibnamefont {Kubica}},
  \bibinfo {author} {\bibfnamefont {S.~T.}\ \bibnamefont {Flammia}},\ and\
  \bibinfo {author} {\bibfnamefont {E.~T.}\ \bibnamefont {Campbell}},\
  }\bibfield  {title} {\bibinfo {title} {Fragile boundaries of tailored surface
  codes},\ }\href@noop {} {\bibfield  {journal} {\bibinfo  {journal} {arXiv
  preprint arXiv:2203.04948}\ } (\bibinfo {year} {2022})}\BibitemShut {NoStop}%
\bibitem [{\citenamefont {Horsman}\ \emph {et~al.}(2012)\citenamefont
  {Horsman}, \citenamefont {Fowler}, \citenamefont {Devitt},\ and\
  \citenamefont {Van~Meter}}]{horsman2012surface}%
  \BibitemOpen
  \bibfield  {author} {\bibinfo {author} {\bibfnamefont {C.}~\bibnamefont
  {Horsman}}, \bibinfo {author} {\bibfnamefont {A.~G.}\ \bibnamefont {Fowler}},
  \bibinfo {author} {\bibfnamefont {S.}~\bibnamefont {Devitt}},\ and\ \bibinfo
  {author} {\bibfnamefont {R.}~\bibnamefont {Van~Meter}},\ }\bibfield  {title}
  {\bibinfo {title} {Surface code quantum computing by lattice surgery},\
  }\href@noop {} {\bibfield  {journal} {\bibinfo  {journal} {New Journal of
  Physics}\ }\textbf {\bibinfo {volume} {14}},\ \bibinfo {pages} {123011}
  (\bibinfo {year} {2012})}\BibitemShut {NoStop}%
\bibitem [{\citenamefont {Chamberland}\ and\ \citenamefont
  {Campbell}(2022{\natexlab{b}})}]{chamberland2022circuit}%
  \BibitemOpen
  \bibfield  {author} {\bibinfo {author} {\bibfnamefont {C.}~\bibnamefont
  {Chamberland}}\ and\ \bibinfo {author} {\bibfnamefont {E.~T.}\ \bibnamefont
  {Campbell}},\ }\bibfield  {title} {\bibinfo {title} {Circuit-level protocol
  and analysis for twist-based lattice surgery},\ }\href@noop {} {\bibfield
  {journal} {\bibinfo  {journal} {Physical Review Research}\ }\textbf {\bibinfo
  {volume} {4}},\ \bibinfo {pages} {023090} (\bibinfo {year}
  {2022}{\natexlab{b}})}\BibitemShut {NoStop}%
\bibitem [{\citenamefont {Fawzi}\ \emph {et~al.}(2018)\citenamefont {Fawzi},
  \citenamefont {Grospellier},\ and\ \citenamefont
  {Leverrier}}]{fawzi2018constant}%
  \BibitemOpen
  \bibfield  {author} {\bibinfo {author} {\bibfnamefont {O.}~\bibnamefont
  {Fawzi}}, \bibinfo {author} {\bibfnamefont {A.}~\bibnamefont {Grospellier}},\
  and\ \bibinfo {author} {\bibfnamefont {A.}~\bibnamefont {Leverrier}},\
  }\bibfield  {title} {\bibinfo {title} {Constant overhead quantum
  fault-tolerance with quantum expander codes},\ }in\ \href@noop {} {\emph
  {\bibinfo {booktitle} {2018 IEEE 59th Annual Symposium on Foundations of
  Computer Science (FOCS)}}}\ (\bibinfo {organization} {IEEE},\ \bibinfo {year}
  {2018})\ pp.\ \bibinfo {pages} {743--754}\BibitemShut {NoStop}%
\bibitem [{\citenamefont {Krinner}\ \emph {et~al.}(2021)\citenamefont
  {Krinner}, \citenamefont {Lacroix}, \citenamefont {Remm}, \citenamefont
  {Di~Paolo}, \citenamefont {Genois}, \citenamefont {Leroux}, \citenamefont
  {Hellings}, \citenamefont {Lazar}, \citenamefont {Swiadek}, \citenamefont
  {Herrmann} \emph {et~al.}}]{krinner2021realizing}%
  \BibitemOpen
  \bibfield  {author} {\bibinfo {author} {\bibfnamefont {S.}~\bibnamefont
  {Krinner}}, \bibinfo {author} {\bibfnamefont {N.}~\bibnamefont {Lacroix}},
  \bibinfo {author} {\bibfnamefont {A.}~\bibnamefont {Remm}}, \bibinfo {author}
  {\bibfnamefont {A.}~\bibnamefont {Di~Paolo}}, \bibinfo {author}
  {\bibfnamefont {E.}~\bibnamefont {Genois}}, \bibinfo {author} {\bibfnamefont
  {C.}~\bibnamefont {Leroux}}, \bibinfo {author} {\bibfnamefont
  {C.}~\bibnamefont {Hellings}}, \bibinfo {author} {\bibfnamefont
  {S.}~\bibnamefont {Lazar}}, \bibinfo {author} {\bibfnamefont
  {F.}~\bibnamefont {Swiadek}}, \bibinfo {author} {\bibfnamefont
  {J.}~\bibnamefont {Herrmann}}, \emph {et~al.},\ }\bibfield  {title} {\bibinfo
  {title} {Realizing repeated quantum error correction in a distance-three
  surface code},\ }\href@noop {} {\bibfield  {journal} {\bibinfo  {journal}
  {arXiv preprint arXiv:2112.03708}\ } (\bibinfo {year} {2021})}\BibitemShut
  {NoStop}%
\bibitem [{\citenamefont {Acharya}\ \emph {et~al.}(2022)\citenamefont
  {Acharya}, \citenamefont {Aleiner}, \citenamefont {Allen}, \citenamefont
  {Andersen}, \citenamefont {Ansmann}, \citenamefont {Arute}, \citenamefont
  {Arya}, \citenamefont {Asfaw}, \citenamefont {Atalaya}, \citenamefont
  {Babbush} \emph {et~al.}}]{acharya2022suppressing}%
  \BibitemOpen
  \bibfield  {author} {\bibinfo {author} {\bibfnamefont {R.}~\bibnamefont
  {Acharya}}, \bibinfo {author} {\bibfnamefont {I.}~\bibnamefont {Aleiner}},
  \bibinfo {author} {\bibfnamefont {R.}~\bibnamefont {Allen}}, \bibinfo
  {author} {\bibfnamefont {T.~I.}\ \bibnamefont {Andersen}}, \bibinfo {author}
  {\bibfnamefont {M.}~\bibnamefont {Ansmann}}, \bibinfo {author} {\bibfnamefont
  {F.}~\bibnamefont {Arute}}, \bibinfo {author} {\bibfnamefont
  {K.}~\bibnamefont {Arya}}, \bibinfo {author} {\bibfnamefont {A.}~\bibnamefont
  {Asfaw}}, \bibinfo {author} {\bibfnamefont {J.}~\bibnamefont {Atalaya}},
  \bibinfo {author} {\bibfnamefont {R.}~\bibnamefont {Babbush}}, \emph
  {et~al.},\ }\bibfield  {title} {\bibinfo {title} {Suppressing quantum errors
  by scaling a surface code logical qubit},\ }\href@noop {} {\bibfield
  {journal} {\bibinfo  {journal} {arXiv preprint arXiv:2207.06431}\ } (\bibinfo
  {year} {2022})}\BibitemShut {NoStop}%
\bibitem [{\citenamefont {Das}\ \emph {et~al.}(2021)\citenamefont {Das},
  \citenamefont {Locharla},\ and\ \citenamefont {Jones}}]{das2021lilliput}%
  \BibitemOpen
  \bibfield  {author} {\bibinfo {author} {\bibfnamefont {P.}~\bibnamefont
  {Das}}, \bibinfo {author} {\bibfnamefont {A.}~\bibnamefont {Locharla}},\ and\
  \bibinfo {author} {\bibfnamefont {C.}~\bibnamefont {Jones}},\ }\bibfield
  {title} {\bibinfo {title} {Lilliput: A lightweight low-latency lookup-table
  based decoder for near-term quantum error correction},\ }\href@noop {}
  {\bibfield  {journal} {\bibinfo  {journal} {arXiv preprint arXiv:2108.06569}\
  } (\bibinfo {year} {2021})}\BibitemShut {NoStop}%
\bibitem [{\citenamefont {{Riverlane Team}}(2022)}]{riverlane2022whitepaper}%
  \BibitemOpen
  \bibfield  {author} {\bibinfo {author} {\bibnamefont {{Riverlane Team}}},\
  }\href
  {https://www.riverlane.com/app/uploads/2022/09/Deltaflow_Decode_Technical_White_Paper_September_2022.pdf}
  {\bibinfo {title} {{Deltaflow.Decode Technical White Paper}}} (\bibinfo
  {year} {2022})\BibitemShut {NoStop}%
\bibitem [{\citenamefont {Fowler}(2015)}]{Fowler2013O1}%
  \BibitemOpen
  \bibfield  {author} {\bibinfo {author} {\bibfnamefont {A.}~\bibnamefont
  {Fowler}},\ }\bibfield  {title} {\bibinfo {title} {Minimum weight perfect
  matching of fault-tolerant topological quantum error correction in average
  $o(1)$ parallel time},\ }\href@noop {} {\bibfield  {journal} {\bibinfo
  {journal} {Quantum Information and Computation}\ }\textbf {\bibinfo {volume}
  {15}},\ \bibinfo {pages} {145} (\bibinfo {year} {2015})}\BibitemShut
  {NoStop}%
\bibitem [{\citenamefont {Anwar}\ \emph {et~al.}(2014)\citenamefont {Anwar},
  \citenamefont {Brown}, \citenamefont {Campbell},\ and\ \citenamefont
  {Browne}}]{anwar2014fast}%
  \BibitemOpen
  \bibfield  {author} {\bibinfo {author} {\bibfnamefont {H.}~\bibnamefont
  {Anwar}}, \bibinfo {author} {\bibfnamefont {B.~J.}\ \bibnamefont {Brown}},
  \bibinfo {author} {\bibfnamefont {E.~T.}\ \bibnamefont {Campbell}},\ and\
  \bibinfo {author} {\bibfnamefont {D.~E.}\ \bibnamefont {Browne}},\ }\bibfield
   {title} {\bibinfo {title} {Fast decoders for qudit topological codes},\
  }\href@noop {} {\bibfield  {journal} {\bibinfo  {journal} {New Journal of
  Physics}\ }\textbf {\bibinfo {volume} {16}},\ \bibinfo {pages} {063038}
  (\bibinfo {year} {2014})}\BibitemShut {NoStop}%
\bibitem [{\citenamefont {Ueno}\ \emph {et~al.}(2021)\citenamefont {Ueno},
  \citenamefont {Kondo}, \citenamefont {Tanaka}, \citenamefont {Suzuki},\ and\
  \citenamefont {Tabuchi}}]{ueno2021qecool}%
  \BibitemOpen
  \bibfield  {author} {\bibinfo {author} {\bibfnamefont {Y.}~\bibnamefont
  {Ueno}}, \bibinfo {author} {\bibfnamefont {M.}~\bibnamefont {Kondo}},
  \bibinfo {author} {\bibfnamefont {M.}~\bibnamefont {Tanaka}}, \bibinfo
  {author} {\bibfnamefont {Y.}~\bibnamefont {Suzuki}},\ and\ \bibinfo {author}
  {\bibfnamefont {Y.}~\bibnamefont {Tabuchi}},\ }\bibfield  {title} {\bibinfo
  {title} {Qecool: On-line quantum error correction with a superconducting
  decoder for surface code},\ }in\ \href@noop {} {\emph {\bibinfo {booktitle}
  {2021 58th ACM/IEEE Design Automation Conference (DAC)}}}\ (\bibinfo
  {organization} {IEEE},\ \bibinfo {year} {2021})\ pp.\ \bibinfo {pages}
  {451--456}\BibitemShut {NoStop}%
\bibitem [{\citenamefont {Meinerz}\ \emph {et~al.}(2022)\citenamefont
  {Meinerz}, \citenamefont {Park},\ and\ \citenamefont
  {Trebst}}]{meinerz2022scalable}%
  \BibitemOpen
  \bibfield  {author} {\bibinfo {author} {\bibfnamefont {K.}~\bibnamefont
  {Meinerz}}, \bibinfo {author} {\bibfnamefont {C.-Y.}\ \bibnamefont {Park}},\
  and\ \bibinfo {author} {\bibfnamefont {S.}~\bibnamefont {Trebst}},\
  }\bibfield  {title} {\bibinfo {title} {Scalable neural decoder for
  topological surface codes},\ }\href@noop {} {\bibfield  {journal} {\bibinfo
  {journal} {Physical Review Letters}\ }\textbf {\bibinfo {volume} {128}},\
  \bibinfo {pages} {080505} (\bibinfo {year} {2022})}\BibitemShut {NoStop}%
\bibitem [{\citenamefont {Paler}\ and\ \citenamefont
  {Fowler}(2022)}]{paler2022pipelined}%
  \BibitemOpen
  \bibfield  {author} {\bibinfo {author} {\bibfnamefont {A.}~\bibnamefont
  {Paler}}\ and\ \bibinfo {author} {\bibfnamefont {A.~G.}\ \bibnamefont
  {Fowler}},\ }\bibfield  {title} {\bibinfo {title} {Pipelined correlated
  minimum weight perfect matching of the surface code},\ }\href@noop {}
  {\bibfield  {journal} {\bibinfo  {journal} {arXiv preprint arXiv:2205.09828}\
  } (\bibinfo {year} {2022})}\BibitemShut {NoStop}%
\bibitem [{\citenamefont {Ueno}\ \emph {et~al.}(2022)\citenamefont {Ueno},
  \citenamefont {Kondo}, \citenamefont {Tanaka}, \citenamefont {Suzuki},\ and\
  \citenamefont {Tabuchi}}]{ueno2022neo}%
  \BibitemOpen
  \bibfield  {author} {\bibinfo {author} {\bibfnamefont {Y.}~\bibnamefont
  {Ueno}}, \bibinfo {author} {\bibfnamefont {M.}~\bibnamefont {Kondo}},
  \bibinfo {author} {\bibfnamefont {M.}~\bibnamefont {Tanaka}}, \bibinfo
  {author} {\bibfnamefont {Y.}~\bibnamefont {Suzuki}},\ and\ \bibinfo {author}
  {\bibfnamefont {Y.}~\bibnamefont {Tabuchi}},\ }\bibfield  {title} {\bibinfo
  {title} {Neo-qec: Neural network enhanced online superconducting decoder for
  surface codes},\ }\href@noop {} {\bibfield  {journal} {\bibinfo  {journal}
  {arXiv preprint arXiv:2208.05758}\ } (\bibinfo {year} {2022})}\BibitemShut
  {NoStop}%
\bibitem [{\citenamefont {Fowler}\ and\ \citenamefont
  {Gidney}(2018)}]{fowler2018low}%
  \BibitemOpen
  \bibfield  {author} {\bibinfo {author} {\bibfnamefont {A.~G.}\ \bibnamefont
  {Fowler}}\ and\ \bibinfo {author} {\bibfnamefont {C.}~\bibnamefont
  {Gidney}},\ }\bibfield  {title} {\bibinfo {title} {Low overhead quantum
  computation using lattice surgery},\ }\href@noop {} {\bibfield  {journal}
  {\bibinfo  {journal} {arXiv preprint arXiv:1808.06709}\ } (\bibinfo {year}
  {2018})}\BibitemShut {NoStop}%
\bibitem [{\citenamefont {Bomb{\'\i}n}(2013)}]{bombin2013introduction}%
  \BibitemOpen
  \bibfield  {author} {\bibinfo {author} {\bibfnamefont {H.}~\bibnamefont
  {Bomb{\'\i}n}},\ }\bibfield  {title} {\bibinfo {title} {An introduction to
  topological quantum codes},\ }\href@noop {} {\bibfield  {journal} {\bibinfo
  {journal} {arXiv preprint arXiv:1311.0277}\ } (\bibinfo {year}
  {2013})}\BibitemShut {NoStop}%
\bibitem [{\citenamefont {Kubica}\ and\ \citenamefont
  {Beverland}(2015)}]{KubicaBeverland15}%
  \BibitemOpen
  \bibfield  {author} {\bibinfo {author} {\bibfnamefont {A.}~\bibnamefont
  {Kubica}}\ and\ \bibinfo {author} {\bibfnamefont {M.~E.}\ \bibnamefont
  {Beverland}},\ }\bibfield  {title} {\bibinfo {title} {Universal transversal
  gates with color codes: A simplified approach},\ }\href
  {https://doi.org/10.1103/PhysRevA.91.032330} {\bibfield  {journal} {\bibinfo
  {journal} {Phys. Rev. A}\ }\textbf {\bibinfo {volume} {91}},\ \bibinfo
  {pages} {032330} (\bibinfo {year} {2015})}\BibitemShut {NoStop}%
\end{thebibliography}%

\end{document}